# Metastable $CoCu_2O_3$ for molecular sensing and catalysis

Matteo D'Andria,[1] Tiago Elias Abi-Ramia Silva,[1] Edoardo Consogno,[1]

Frank Krumeich[2] and Andreas T. Güntner[1,*]

[1] *Human-centered Sensing Laboratory, Department of Mechanical and Process Engineering, ETH Zurich, CH-8092 Zurich, Switzerland*
[2] *Laboratory of Inorganic Chemistry, Department of Chemistry and Applied Biosciences, ETH Zurich, CH-8093 Zurich, Switzerland*

* corresponding author: andregue@ethz.ch

# Abstract

Metastable nanostructures are kinetically trapped in local energy minima featuring intriguing surface and material properties. To unleash their potential, there is a need for non-equilibrium processes capable of stabilizing a large range of crystal phases outside thermodynamic equilibrium conditions by closely and flexibly controlling atomic reactant composition, spatial temperature distribution and residence time. Here, we demonstrate the capture of metastable pseudo-binary metal oxides at room temperature with scalable combustion-aerosol processes. By a combination of X-ray diffraction, electron microscopy and on-line flame characterization, we investigate the occurrence of metastable $CoCu_2O_3$ with controlled crystal size (4 – 16 nm) over thermodynamically stable CuO and $Co_3O_4$. Immediate practical impact is demonstrated by exceptional sensing and catalytic performance for air pollutant detection (e.g., 15 parts-per-billion benzene). This approach can be extended to various binary, ternary and high entropy oxides with even more components to access novel materials also promising for actuators, energy storage or solar cells.

# 1 Introduction

Advancements in nanotechnology are mostly achieved by material innovation with crystalline materials (e.g., microporous COFs[1]/MOFs,[2] MXenes,[3] and perovskites[4]) that are thermodynamically stable at room temperature, with direct impact for applications including molecular sensing,[5] catalysis,[6] batteries/energy storage[3] and solar cells.[7] Mostly neglected are metastable phases, that can be found extensively in binary (e.g., W-O,[8,9] Fe-O[10]), ternary (Co-Cu-O,[11] In-Sn-O,[12] Co-Mo-O,[13] W-Mo-O[14]), or high entropy oxide systems with even more components.[15] Kinetically trapped in local minima on complex free energy hypersurfaces, they can offer unique properties. This is associated, for instance, with higher surface energies due to their less stable atomic configuration,[16] which may improve molecular interaction, fostering surface-active processes such as heterogeneous catalysis[17] and molecular sensing.[18]

To unlock the potential of metastable materials, non-equilibrium processes need to be explored for their capacity to access a variety of bulk structures across the high- and low-temperature phases detailed in thermodynamic equilibrium diagrams. Key necessities are control over: (i) ionic/atomic composition of reactants, (ii) spatial temperature distribution, (iii) material residence time and (iv) its rapid quenching to be able to form and "freeze" the metastable phase in a non-equilibrium arrangement (e.g., at room temperature). Ideally, there is high flexibility to mix various reactants and coverage over a large temperature range to access metastable phases in various materials systems.[8-15] Furthermore, flexibility in residence time within the stability region of these metastable phases will enable close control over crystal size, a key property as it determines, e.g., sensitivity in molecular sensing.[19]

Ultra-Fast Joule Heating (UFJH)[20] has recently drawn attention for this purpose. Therein, the temperature can be controlled within a large range (1200 – 2000 °C) by modulating electrical power through the nanoparticles, and the quenching is as fast as $10^4$ K $s^{-1}$.[21] Yet, UFJH provides insufficient residence time at high temperatures (e.g., 10 ms[20]) to nucleate and grow the metastable phase, with few exceptions such as $\alpha$-$MoC_{1-x}$/$\eta$-$MoC_{1-x}$[22] and turbostratic BCN,[23] next to limited scalability and throughput (batch process). As a result, thermodynamically stable crystalline materials with very high concentrations of defects[24] (e.g., twin boundaries, vacancies) and stresses are usually obtained, rather than metastable phases with different symmetries. Such prepared materials can yield unprecedented properties, e.g, Si-nanowires dispersed on reduced graphene oxide are applicable for Li-ion batteries.[25]

Combustion aerosol processes (e.g., flame spray pyrolysis[26]) provide access to the necessary temperature range (1000 – 3000 °C[27]) to capture high-temperature crystal phases with particle nuclei experiencing enough residence time (e.g., hundreds of milliseconds[27]) to grow before being rapidly quenched. Further, it is thereby possible to control the particle[28] and aggregate size thanks to the rapid attainment of the self-preserving size distribution.[29] As a result, combustion processes seem well suited to manufacture metastable phases, despite their primary application for the synthesis of thermodynamically stable materials (e.g., cubic $Co_3O_4$[30]). Such continuous processes are proven to be scalable,[27] and have already led to commercialized products[31] (e.g., handheld sensor devices[32,33]).

Here, we examine a combustion-aerosol process design to systematically capture a high-temperature phase and its meta-stabilization at room temperature, as demonstrated with $CoCu_2O_3$. We have chosen the CuO-$Co_3O_4$ pseudo-binary system because of the rich presence of mixed oxide phases (e.g., spinel $Cu_xCo_{3-x}O_4$[34,35], rock-salt $Cu_yCo_{1-y}O$[36]) thanks to the high Cu- and Co-miscibility due to similar ionic radii.[37] By a combination of ex/in-situ X-ray diffraction, electron microscopy and on-line flame characterization, we investigate $CoCu_2O_3$ nanocrystals synthesis and thermal stability. We establish general thermodynamics-process relationships to control product phase composition as a function of the flame-aerosol engineering conditions that affect ion concentration, temperature profile and high-temperature residence time. To harness the immediate practical potential of $CoCu_2O_3$, we systematically assess its electronic, optical, chemoresistive and catalytic properties.

# 2 Results

## 2.1 Combustion-aerosol synthesis of metastable $CoCu_2O_3$

Figure 1a shows a schematic illustration of the flame spray pyrolysis (FSP) process: the liquid precursor containing the Cu and Co metals is fed through a nozzle, dispersed into a fine spray, and ignited by a premixed flamelet under atmospheric conditions.[5] The as-produced Cu/Co-enriched vapor forms particle nuclei that grow by condensation and order in crystalline structures by solid-state diffusion.[38] Meanwhile, aerosol coagulation through collisions driven by the high flame temperatures results in aggregates chemically bonded through sinter necks, before being rapidly quenched to low temperatures.[39] To demonstrate control over a large temperature range, we varied the precursor (P, mL $min^{-1}$) to dispersion (D, L $min^{-1}$) ratio (P/D), as shown in Figure 1a for P/D = 2/8, 5/5 and 9/3. Higher P/D ratios yield longer flames and higher temperatures, for instance, 1900 °C (9/3) compared to 400 °C (2/8) at 12 cm above the burner due to the increased energy release rate, which drastically alters the aerosol residence time and temperature history.

The X-ray diffraction (XRD) patterns of as-prepared nanoparticles are shown in Figure 1b. Pure $Co_3O_4$ and CuO form the cubic[40] and monoclinic[41] phases, respectively. At Co:Cu = 1:2, metastable $CoCu_2O_3$ is obtained in powders at room temperature for P/D of 2/8 and 5/5, as evidenced by new peaks at $2\theta \approx 34.0$, 40.2 and 45.7°. Rietveld refinement yields lattice parameters close to reference[42] values (Figure S1). According to the CuO-$Co_3O_4$ phase diagram,[11] this phase is thermodynamically stable between 915 – 1070 °C, in line with our observation (Figure S2) following a different protocol. Remarkably, FSP is capable to directly access and stabilize $CoCu_2O_3$ at room temperature. In contrast, previous methods formed it only from a reactive precursor mixture (CoO/$Co_2O_3$), and after lengthy high-temperature treatments and crystallization from a potassium-fluoride-flux melt.[42]

The powders appear crystalline, as indicated by faceted surfaces and clearly visible lattice fringes in transmission electron microscopy (TEM). Figure 1c shows such a $CoCu_2O_3$ nanocrystal oriented along the $[10\bar{3}]$ direction featuring horizontal and vertical interlayer spacings ($d_{lattice}$) of 2.05 and 2.2 Å associated to the (020) and (301) planes, respectively, in agreement with XRD (Figure 1b). This is further supported by selected area electron diffraction (SAED), where multiple matching ring-like patterns are visible (Figure 1e). Note that corresponding TEM and SAED with identification of the *201* facet of $CoCu_2O_3$ for 2/8 powders are provided in Figure S3.

Only for hotter flames (e.g., P/D = 7/3), XRD peaks associated with cubic $Co_3O_4$ appear at $2\theta \approx 31.1$, 59.1 and 65.2° (Figure 1b) which become more pronounced at higher P/D = 9/3, accounting for an estimated $Co_3O_4$ amount of ~ 30 wt% (Figure S4). Therein, $d_{lattice}$ = 2.42 Å (Figure 1d) matches well the $Co_3O_4$ – (311) plane. In the electron diffraction pattern (Figure 1f), the two innermost weak rings correspond to the reflections *111* and *220* of $Co_3O_4$ (see Figure S5 for magnified pattern), which is also identified through a bright spot indexed to *311* (inset). The elemental map derived from energy dispersive X-ray spectroscopy (EDXS) in Figure S6 confirms the presence of Cu-free domains corresponding to $Co_3O_4$ crystallites. This segregation is attributed to the sufficiently long residence time of $CoCu_2O_3$ nanocrystals in the flame (Figure 1a, circles) at temperatures below their thermodynamic stability range (< 915 °C),[11] so that it partly recrystallizes into the energetically more favorable (at these conditions) cubic $Co_3O_4$ structure. Note that the overall composition is also a critical parameter (Figure S7), as pure metastable $CoCu_2O_3$ is only obtained at a ratio Co:Cu = 1:2, while mixtures with CuO, $Co_3O_4$ and/or $Cu_xCo_{3-x}O_4$ are formed when deviating from that stoichiometry.

Finally, Figure 1g shows the $CoCu_2O_3$ crystal ($d_{XRD}$) size that increases from ~ 5 nm to 15 nm when increasing the P/D ratio from 2/8 to 9/3, in agreement with TEM observations (Figure 1c,d and Figure S3, S6). This corroborates their longer high-temperature residence time at higher P/D. The obtained particles are monocrystalline, as $d_{XRD}$ and BET-equivalent ($d_{BET}$) sizes are similar, in agreement with TEM (Figure 1c,d) and as had been observed before for flame-made $CuO_x/Co_3O_4$.[30]

## 2.2 Thermal stability of metastable $CoCu_2O_3$

To probe the thermal stability of $CoCu_2O_3$, 5-hour annealing treatments at increasing temperatures up to 600 °C were performed. XRD patterns (Figure 2a) reveal that $CoCu_2O_3$ is preserved until 300 °C, when small features corresponding to cubic $Cu_{0.95}Co_{2.05}O_4$ emerge at 2θ ≈ 30.6 and 64.6°. Interestingly, peaks of $Cu_xCo_{3-x}O_4$ become more pronounced at higher annealing temperatures, accompanied by a gradual change in stoichiometry from $x$ = 0.95 to 0.72 at 500 °C. These results suggest the progressive migration of Cu out of the crystal and formation of cubic spinel, as confirmed by the simultaneous appearance of increasingly sharp

CuO peaks at 2θ ≈ 36.6 and 38.8°. Finally, the segregation of Cu- and Co-containing crystallites is completed at 600 °C, when a mixture of CuO and $Co_3O_4$ is observed, in agreement with the CuO-$Co_3O_4$ equilibrium diagram.[11]

This thermal behaviour is confirmed by elemental maps of powders annealed at 200 and 600 °C in Figure 2d and e, respectively. Most importantly, the distribution of Co (blue) and Cu (green) appears rather homogeneous throughout the various $CoCu_2O_3$ aggregated nanoparticles (Figure 2d) and the EDX-derived elemental composition of ca. 64.5 at% Cu from Figure 2f is close to the nominal Co:Cu atomic ratio (i.e., 1:2). However, some green spots indicate few Cu-enriched domains. In contrast, when annealed at 600 °C, Co-rich (area II) and Cu-rich (III) areas are clearly distinguished, as confirmed by their EDX spectra in Figure 2g and h, respectively, evidencing the segregation of $CoCu_2O_3$ into $Co_3O_4$ and CuO crystals at high temperature, in line with the diffractograms in Figure 2a.

The corresponding particle sizes determined by BET and crystal sizes for $CoCu_2O_3$, CuO and $Co_3O_4$ are shown in Figure 2b. We note that $d_{BET}$ and $d_{XRD}$ remain stable (~ 9 nm) up to 300 °C, defining the thermally stable range for metastable $CoCu_2O_3$ nanocrystals. Thereafter, particles grow rapidly reaching a particle size of ~ 70 nm at 600 °C, while CuO and $Co_3O_4$'s crystal sizes both reach ~ 38 nm. Meanwhile, the particles change from monocrystalline ($d_{BET} \approx d_{XRD}$, up to 300 °C) to polycrystalline ($d_{BET} > d_{XRD}$), which might be attributed to thermally activated sintering of the necks[43] between as-produced primary particles, as observed by TEM (Figure S8).

To evaluate also the $CoCu_2O_3$ phase stability over longer periods, we placed our powders in a non-ambient chamber and recorded the XRD patterns daily when kept at 200 °C for 9 days (Figure 2c). Importantly, the main $CoCu_2O_3$ peaks (i.e., 2θ ≈ 36.0, 36.58, 40.23 and 45.72°) are mostly preserved for the entire duration. We only note the emergence of a broad peak at 2θ ≈ 65°, which could indicate traces of $Cu_xCo_{3-x}O_4$ (Figure 2a). However, if higher temperatures are needed in certain applications, the thermal stability may be further enhanced by defect incorporation through foreign element doping, as demonstrated with Cr[44] or Si[45] for $\varepsilon$-$WO_3$.

## 2.3 Optical, electronic and chemoresistive properties of $CoCu_2O_3$

For assessing the electronic and optical properties of the metastable $CoCu_2O_3$ nanocrystals, we selected particles produced at P/D = 5/5 after annealing at 200 °C for 5 hours to ensure high phase purity (Figures 1b and 2a) and contaminant-free surfaces. As shown in the UV-Vis-NIR spectra (Figure 3a), $CoCu_2O_3$ has a feature at 815 nm assigned to *d-d* transitions among non-degenerate[46] and distorted[47] (Jahn-Teller effect) $t_{2g}$ and $e_g$ orbitals of $Cu^{2+}$, as similarly observed for CuO, though slightly shifted towards ~ 800 nm. The rest of the spectrum is dominated by a steep increase below 250 nm, due to the ligand-to-metal charge transfer (LMCT) $O^{2-}$ (2p) → $Cu^{2+}$ (3d).[48] This is significantly different from $Co_3O_4$, which presents distinct peaks[49] associated with the LMCTs (i) $O^{2-}$ (2p) → $Co^{3+}$ ($e_g$) at 720 nm and (ii) $O^{2-}$ (2p) → $Co^{2+}$ ($t_2$) at 405 nm.

To establish a quantitative footing, we estimated the direct bandgap energy ($E_g$) from the optical absorbance (see Tauc plots in Figure S9). Importantly, the similarity between $E_g$ of $CoCu_2O_3$ (i.e., 1.34 eV) and the energy of the *d-d* transitions of $Cu^{2+}$ indicates a strong hybridization between *d* orbitals and the states close to the Fermi level ($E_F$).[50] Therefore, the slight shift observed in Figure 3a could be due to the *d*-band center ($\varepsilon_d$) being closer to $E_F$ in $CoCu_2O_3$,[51] a promising feature of metastable $CoCu_2O_3$ that can foster analyte adsorption[52] and catalytic performance.[53] Further optical bandgaps at different Co:Cu ratios are given in Figure S10. Interestingly, the optical $E_g$ of 1.34 eV (at room temperature) of $CoCu_2O_3$ is in excellent agreement with *ab initio* electronic structure calculations[54] (1.25 eV), as well as with the thermal $E_g$ of 1.33 eV (at 0 K,[55] derived from temperature-dependent resistance measurements shown in Figure S11), possibly due to low thermal expansion[56] or weak electron-phonon coupling.[57]

Next, we investigated the potential of $CoCu_2O_3$ for chemoresistive sensing to demonstrate immediate practical impact. Figure 3c shows the resistance of $Co_3O_4$, $CoCu_2O_3$ and CuO porous films at 160 °C in $N_2$, dry air, and air with 50% relative humidity (RH). In $N_2$, the resistance of $CoCu_2O_3$ (57 kΩ) is lower than those of similarly deposited $Co_3O_4$ (148 kΩ) and CuO (20 MΩ), which is desirable for sensor integration in monolithic systems.[58] Further, $CoCu_2O_3$ shows p-type behavior, as its resistance decreases when exposed to air (Figure 3c), likely due to O-related species that adsorb on the surface capturing electrons and building up a hole accumulation layer, as described for CuO and $Co_3O_4$.[59] Remarkably, the resistance of $CoCu_2O_3$ hardly changes between dry and 50% RH (see inset in Figure 3c), suggesting high robustness, a favorable property of gas sensors that usually requires challenging surface engineering (e.g., $CeO_2$ cluster formation on $In_2O_3$[60]). When exploited as a molecular sensor for the critical air pollutant xylene,[61] even the lowest levels down to 12 ppb are detected (Figure 3d). In fact, metastable $CoCu_2O_3$ shows clearly distinguishable responses between 12 – 300 ppb with exceptional signal-to-noise ratio (≥ 40), and full recovery after each exposure within 5 – 10 minutes at a moderate temperature of 200 °C. This detection limit is about an order of magnitude lower than what has been achieved yet by typical state-of-the-art sensors, e.g., noble-metal-doped $WO_3$.[62]

### 2.4 Catalytic oxidation and surface acidity

Catalytic behaviour of metastable $CoCu_2O_3$ was evaluated for the oxidation of molecules from various chemical classes including ketones, alcohols, aromatic compounds and terpenes. Figure 4a shows the conversion of acetone, methanol, xylene, isoprene, ethanol, toluene and benzene in the temperature range of 20 – 200 °C at 50% RH. Remarkably, ethanol starts to be converted at 30 °C, demonstrating the high catalytic activity of $CoCu_2O_3$. The conversion onset of acetone and methanol occurs at ~ 80 °C. Most importantly, benzene reacts on $CoCu_2O_3$ only above 180 °C, clearly differentiating it from xylene and toluene, which are fully converted already at 140 and 160 °C, respectively, despite their chemically similar structure with one aromatic ring. This can be exploited for highly selective benzene detection,[63] as shown below. In contrast, CuO and $Co_3O_4$ display different oxidation kinetics (Figure S12,13), attributed to a wide range of activation energies ($E_a$) and surface chemistries (Figure S14).

The higher reactivity of $CoCu_2O_3$ to xylene and toluene over benzene should be associated with Lewis-acidic surface sites,[64] that better interact with methyl-substituted, electron-rich,[65] aromatic rings for C=C bond cleavage and ring opening. Therefore, we investigated surface acidity by diffuse reflectance infrared Fourier transform spectroscopy (DRIFTS) of adsorbed pyridine (Py).[66] Therein, Py is adsorbed at room temperature, followed by purging under He at increasing temperatures up to 200 °C. As shown in Figure 4b, several features occur in the $v$(CCN) ring-breathing region, which are associated[67,68] with the stretching modes of Py coordinated with Lewis (LPy) and Brønsted (BPy) acid sites, as well as Py bonded to acidic H-atom of hydroxyl groups (HPy). Interestingly, $v_{19a}$ of both BPy and LPy, as well as the $v_{19b}$ of LPy, slightly shift to higher energies (1474 $cm^{-1}$ and 1437 $cm^{-1}$, respectively) on $CoCu_2O_3$ compared to $Co_3O_4$ and CuO, suggesting enhanced acid-site strength.[69] These include Lewis sites that are needed to convert methyl-substituted aromatic rings preferentially over benzene (Figure 4a), similar to combustion-made $WO_3$,[70] which rendered it suitable for selective benzene sensing.[63] Further, $CoCu_2O_3$ has the highest Brønsted-to-Lewis ratio, based on the $v_{19b}$ – intensities around 1540 $cm^{-1}$ (BPy) and 1430 – 1440 $cm^{-1}$ (LPy), which possibly explains the removal of acetone below 100 °C in Figure 4a, compared to ~ 150 °C for CuO and $Co_3O_4$ (Figure S12). In fact, the Brønsted-to-Lewis ratio describes the reactivity patterns of typical Brønsted- and Lewis-coordinated molecules such as acetone[71] and methanol,[72] respectively (Figure S15).

### 2.5 Exclusive benzene detection by $CoCu_2O_3$-filtered chemoresistor

Benzene is a toxic and carcinogenic[73] pollutant that needs to be monitored in critical industries (e.g., rubber[74]), housing,[75] and even during space missions.[76] Its selective detection, particularly in mixtures of other aromatic compounds (e.g., toluene, xylene), is a long-standing challenge due to very low exposure limits (e.g., 100 ppb in US[77] or 50 ppb in EU[78]). Indeed, when testing a widely applied Pd/$SnO_2$ sensor[79] at 350 °C and 50% RH against 1 ppm of benzene and 12 other organic and inorganic molecules, rather similar responses are obtained (Figure 4c). Note that the XRD pattern of Pd/$SnO_2$ is shown in Figure S16, with consistent phase and size as in previous reports.[80]

Most importantly, with a $CoCu_2O_3$ nanoparticle-bed filter operating at 170 °C (Figure 4d), benzene detection remains intact (response of 1.35 vs. 1.37), while the sensor hardly responds anymore ($\leq$ 0.06) to all the other compounds due to their catalytic conversion[81] to sensor-inert species (Figure 4a). As a result, $CoCu_2O_3$ turns Pd/$SnO_2$ into an efficient benzene-selective sensor, with a response ratio of benzene vs. confounder $\geq$ 22. This value is superior to other benzene sensors, for instance, $SnO_2$ with a Rh/$TiO_2$ overlayer[82] (selectivities of ~ 5), or other overlayer-based chemiresistors that allowed exclusive aromatics detection only when deployed in sensor arrays.[83] This composite device is also better than $WO_3$-screened Pd/$SnO_2$ that partially converted benzene (26%) and required higher (240 °C) operational temperature, despite the lower gas weight hourly space velocity of 0.11 vs. 0.56 $mL_{analyte}$ $h^{-1}$ $g_{cat}^{-1}$ used here.

To further challenge our detector, we tested its performance in gas mixtures. Figure 4e shows the response transients to decreasing benzene concentrations of 1000, 500, 300, 100, 50 and 15 ppb, as well as the same with co-present 2000 ppb (each) of toluene and xylene, and even up to 1000 ppb (each) of five different interferants, i.e., toluene, xylene, methanol, ethanol and acetone. Remarkably, the detector is hardly affected by other molecules, despite them being present at two orders-of-magnitude higher concentrations (i.e., at 50 ppb benzene), where steady-state response changes from 0.19 to 0.24 (Figure 4f). The detector is also highly robust to humidity variation (40 – 90%), as shown in Figure 4g.

# 3 Discussion

We rationally devised a non-equilibrium combustion process to synthesize nanocrystals of $CoCu_2O_3$ and stabilize it at ambient conditions, a phase usually observed between 915 – 1070 °C. By in-depth material and flame characterization, we found that flame conditions, precursor composition and high-temperature residence time need to be carefully controlled to avoid segregation into $Co_3O_4$ and CuO crystals. These metastable $CoCu_2O_3$ nanocrystals showed p-type semiconductive properties with a lower bandgap than CuO and $Co_3O_4$, and could be exploited for low-ppb molecular sensing. We identified the acidic surface features of $CoCu_2O_3$ that enabled its use as a catalytic filter for exclusive benzene sensing, superior to state-of-the-art detectors. From a broader perspective, this work demonstrates the rigorous engineering of a non-equilibrium process to access and stabilize high-temperature phases inaccessible by conventional techniques to unlock the potential of metastable materials. As demonstrated, these feature unique catalytic, electronic and sensing properties with immediate practical impact that should open new avenues for other fields of application (e.g., chemical waste valorization, batteries, energy storage).

# 4 Methods

## 4.1 Nanoparticle production

CuO-$Co_3O_4$ nanoparticles were prepared by FSP, with a reactor design detailed elsewhere.[28] To prepare the precursor,[30] we mixed cobalt(II) 2-ethylhexanoate (65 wt% in mineral spirits, Sigma Aldrich, Switzerland) and Deca Copper 8 (Borchers, Germany), as dictated by the final Cu content. This mixture was dissolved in pure xylene (mixture of isomers, VWR Chemicals, Switzerland) to obtain a total metal (Co + Cu) molarity of 0.2 mol $L^{-1}$. Thereafter, the precursor was fed through a capillary and dispersed by $O_2$ (pressure drop of 1.6 bar) to form a fine spray. The precursor (P, mL $min^{-1}$) and dispersion (D, L $min^{-1}$) flow rates were varied as specified in the text. The spray was ignited and sustained by a pilot flame of premixed $CH_4$ (at 1.2 L $min^{-1}$, Methane 2.5, PanGas, Switzerland) and $O_2$ (at 3.2 L $min^{-1}$, Pangas, Switzerland). Additionally, 5 L $min^{-1}$ $O_2$ sheath flow was supplied to shield the flame and ensure excess oxidant. The nanoparticles were deposited for 15 min onto water-cooled glass fiber filters (257 mm diameter, GF6, Hahnemühle Fineart, Germany) at a height above the burner of 57 cm aided by a vacuum pump (Seco SV 1025 C, Busch, Switzerland). The particles were carefully removed from the filter with a spatula and the obtained powders were sieved with a 250 µm mesh. For annealing, the powders were thermally stabilized for 5 hours in an oven (CWF 1300, Carbolite Gero, Germany) under ambient air. The flame temperature was measured with an R-type thermocouple (1-mm bead-diameter, Intertechno-Firag AG) in the center of the flame at 5 – 25 cm above the burner. The such obtained temperature was corrected for radiative heat transfer, following the literature.[84]

## 4.2 Material characterization

XRD patterns of powders were acquired with a Bruker D2 Phaser (USA) operated at 30 kV and 10 mA, at 2θ (Cu $K_\alpha$ radiation, $\lambda$ = 1.5406 Å) between 15° and 75°, with scanning step size of 0.01° and a scanning time of 2.3 seconds per step. Crystal phases were identified by comparison of obtained patterns to the structural parameters of cubic $Co_3O_4$ (PDF 42-1467), monoclinic CuO (PDF 72-0629), orthorhombic $CoCu_2O_3$ (PDF 76-0442), cubic $Cu_{0.95}Co_{2.05}O_4$ (PDF 78-2177), cubic $Cu_{0.92}Co_{2.08}O_4$ (PDF 37-0878), cubic $Cu_{0.72}Co_{2.28}O_4$ (PDF 77-0241) and tetragonal $SnO_2$ (PDF 41-1445). To identify the stoichiometry of $Cu_xCo_{3-x}O_4$ ($x \leq 1$), XRD patterns were aligned with tin telluride (SnTe 99.999%, Sigma Aldrich, Switzerland) as crystalline internal standard.[85] Therefore, the powders were mixed (4:1 by weight) with SnTe in a mortar and the as-recorded XRD patterns were aligned to the reference peaks of cubic SnTe (PDF 46-1210). The crystal size ($d_{XRD}$) was evaluated applying by Rietveld's fundamental parameter method,[86] as implemented in the Topas 4.2 (Bruker) software. For the phase stability test over long periods, XRD patterns were obtained in situ by a Bruker AXS D8 Advance diffractometer equipped with a high-temperature non-ambient chamber (HTK 1200N, Anton Paar, Austria), operated at 40 kV and 40 mA at 2θ (Cu $K_\alpha$ radiation) between 30° and 90°. All diffractograms were corrected for Cu $K_{\alpha 2}$ emission and background. In situ XRD at high temperatures was measured at the European Synchrotron Radiation Facility (Beamline BM01). The powder sample was loaded into a glass capillary (0.5 mm diameter) which was mounted on a holder, and heated with a custom-made furnace at a ramp rate of 60 °C $min^{-1}$. XRD patterns were recorded in transmission on a PILATUS@SNBL diffractometer, with continuous 6-seconds exposures at $\lambda$ = 0.72325 Å. The 2D diffraction data from the Pilatus 2 M detector were processed using the SNBL Toolbox and BUBBLE software.[87] Refence peaks of $CoCu_2O_3$, CuO and $Co_3O_4$ were identified with the same PDF accounting for the different source.

$N_2$-sorption isotherms of powders were recorded on a Tristar II Plus (Micromeritics, USA). The specific surface area (SSA) of powders was determined according to Brunauer-Emmet-Teller (BET) theory. Prior to measurement, samples were degassed for 1.5 hours at 120 °C under $N_2$ to remove any adsorbate. The densities ($\rho$) of 6.31 g $cm^{-3}$, 6.48 g $cm^{-3}$, 6.1 g $cm^{-3}$ and 6.13 g $cm^{-3}$ were used for CuO, $CoCu_2O_3$, $Cu_xCo_{3-x}O_4$ and $Co_3O_4$, respectively. Therein, surface equivalent diameters ($d_{\mathrm{BET}}$) were determined according to:

$$d_{BET} = \frac{6}{SSA \cdot \rho}$$

The powders' UV-Vis-NIR spectra were recorded on a Cary 5000 UV-Vis-NIR spectrophotometer (Agilent, USA). High-purity UV-Vis-NIR grade $BaSO_4$ (Sigma Aldrich, Switzerland) was used for baseline correction. Ca. 8 mg of powder were mixed with 72 mg $BaSO_4$ in a mortar to achieve a homogeneous mixture. This was placed in a reaction cell within a Praying Mantis diffuse reflection accessory (both Harrick Scientific, USA) and a flow of 25 mL $min^{-1}$ $N_2$ was applied. Therein, diffuse reflectance spectra were recorded at room temperature, between 2000 nm and 180 nm and converted to Kubelka Munk (KM), $F(R_\infty)$, units:

$$F(R_\infty) = \frac{(1 - R_\infty)^2}{2R_\infty}$$

The KM function was then used to estimate the optical bandgap energy ($E_g$) with Tauc's method ($\gamma$ = 1/2 for direct transitions[88]):

$$(F(R_\infty)\cdot h\nu)^{1/\gamma} = B\cdot\left(h\nu - E_g\right)$$

The thermal $E_g$ was obtained from the temperature-dependence of the electrical resistance (measured as described in section 4.4) of chemoresistive films, as described in the literature.[89]

Adsorption of pyridine for surface-acidity characterization[66] was investigated by DRIFTS using a Vertex 70v spectrometer (Bruker Optics, USA) equipped with a liquid-nitrogen-cooled mercury cadmium telluride (MCT) detector. Therein, ca. 10 mg of powder were homogeneously dispersed with 90 mg of non-IR active FTIR grade KBr (Sigma Aldrich) that was placed in a high-temperature reaction cell within the Praying Mantis diffuse reflection accessory. Initially, the sample was heated to 100 °C for 15 min under a 30 mL $min^{-1}$ He flow (Pangas, 5.0), and then cooled down to room temperature under the same atmosphere. Thereafter, He was bubbled through technical pyridine (Sigma Aldrich) and the resulting pyridine-rich gas was fed to the reaction cell for 45 min. This was followed by 30-min pyridine desorption under a pyridine-free He-purge at room temperature first, and at increasing temperatures up to 200 °C. Meanwhile, DRIFT spectra were recorded and converted to KM units by averaging 150 scans in the range between 700 and 4000 $cm^{-1}$ at 4 $cm^{-1}$ resolution. Spectral assignments are done following the literature.[66,67]

### 4.3 Electron microscopy

For (scanning) transmission electron microscopy ((S)TEM) investigation, the material was dispersed in ethanol and a few drops of the suspension were deposited onto a perforated carbon foil supported on a copper grid. After evaporation of the ethanol, the grid was mounted on the single tilt holder of the microscope. TEM investigations were performed on a double-corrected microscope JEM-ARM300F (GrandARM, JEOL), operated at an acceleration potential of $U_{acc}$ = 300 kV (electron gun: cold-field emitter; $\Delta E \approx 0.35$ eV). The high-resolution capability of this microscope in both TEM and STEM (ca. 80 pm) is due to corrector systems that compensate the spherical aberration of the TEM objective lens and the probe-forming lens, respectively. TEM images and SAED pattern were recorded with a CMOS Camera (Gatan OneView, 4k x 4k pixels). Annular dark field (ADF) STEM and analytical investigations were performed on a JEM-F200 (JEOL) microscope, operated at $U_{acc}$ = 200 kV (cold field emitter). Elemental mappings based on EDXS (JEOL) were measured spotwisely with two SSD detectors attached to the microscope column.

### 4.4 Chemoresistive characterization

For chemoresistive testing of CuO, $Co_3O_4$, $CoCu_2O_3$ and Pd/$SnO_2$, porous films were obtained by depositing particles directly from the aerosol by thermophoresis[90] for 4 min onto water-cooled sensor substrates (electrode type #103, Electronic Design Center, Case Western University, USA) at 20 cm height above the burner. The substrates were made of $Al_2O_3$ with interdigitated Pt electrodes (spacing of 250 µm) on the front and a meander-shaped Pt heater on the back. The precursor for 1 mol% Pd/$SnO_2$ was taken from the literature.[79] The sensors were mounted onto Macor holders and placed in a PTFE chamber. The sensing film was heated by applying a constant voltage to the heater. The temperature was continuously monitored with a multimeter (2700, Keithley) by using the same Pt heater as the resistance temperature detector. The chamber was connected to a gas mixing set-up described elsewhere.[91] Briefly, hydrocarbon-free synthetic air (PanGas, $C_nH_m$ and $NO_x$ < 100 ppb) was used as a carrier gas and the analytes from certified gas standards were admixed by calibrated mass flow controllers (Bronkhorst) to obtain the desired gas mixture composition. The calibrated and certified gas standards (all Pangas) used in this study were: acetone (15.1 ppm in dry synthetic air), xylene (15 ppm in dry synthetic air), toluene (9.4 ppm in dry synthetic air), benzene (15 ppm in dry synthetic air), ethanol (14.8 ppm in dry synthetic air), CO (506 ppm in dry synthetic air), isoprene (20 ppm in dry synthetic air), methanol (14.3 ppm in dry synthetic air), $H_2$ (50 ppm in dry synthetic air), NO (10 ppm in $N_2$), $NO_2$ (22.8 ppm in dry synthetic air), formaldehyde (10 ppm in $N_2$) and $N_2O$ (10 ppm in $N_2$). Humid air was generated by bubbling dry synthetic air at room temperature through a bubbler filled with deionized water that was admixed to the analyte-containing gas stream. The RH at room temperature was varied between 0 – 90%, as checked with a SHT2x sensor (Sensirion AG, Switzerland), and total flow was kept at 300 mL $min^{-1}$. The ohmic resistance of the sensing film was measured continuously between the interdigitated Pt electrodes with a multimeter (2700, Keithley). The chemoresistive response ($S$) to reducing gases was defined as the normalized resistance variation. For p-type sensors (i.e., CuO, $Co_3O_4$ and $CoCu_2O_3$), this was evaluated as:

$$S = \frac{R_g - R_a}{R_a}$$

while for Pd/$SnO_2$ (n-type) this was defined as:

$$S = \frac{R_a - R_g}{R_g}$$

where $R_g$ and $R_a$ are the resistances of the sensing film under gas exposure and in clean air, respectively.

### 4.5 Catalytic characterization

For the catalytic activity tests, 16 mg ($m_{cat}$) of powder were filled into a glass tube and fixed tightly as a packed bed with quartz wool, as described elsewhere.[92] The quartz tube was placed in a horizontal oven (Carbolite ESZ 12/450, Germany) and connected to the gas mixing setup described above at a total flow rate of 150 mL $min^{-1}$ and 50% RH. The packed bed was heated at 10 °C $min^{-1}$ and a dwell time of 20 min at each temperature was applied, before feeding the analyte-containing mixture. The inlet analyte concentration was 1 ppm and the off-gas was analyzed using a proton transfer reaction time-of-flight mass spectrometer (Ionicon PTR-ToF-MS 1000, Innsbruck, Austria). $H_3O^+$ was used as an ion source and the PTR-ToF-MS was operated with a drift voltage, temperature and pressure of 600 V, 60 °C and 2.3 mbar, respectively. The reduced electric field (E/N) in the drift tube was 130 Td. With deployed catalyst mass and flow conditions, the gas weight hourly space velocity was kept at 0.56 $mL_{analyte}$ $h^{-1}$ $g_{cat}^{-1}$. The analyte concentrations were evaluated at *m/z* values of 33.04 (methanol[93]), 47.05 (ethanol[94]), 59.05 (acetone[95]), 107.16 (*m*-xylene[93]), 93.14 (toluene[93]), 79.05 (benzene[96]) and 69.07 (isoprene[93]). The mass spectrometer was calibrated with 5 points in the range of 0 – 1000 ppb with the aforementioned gas standards for each analyte. The catalytic conversion ($\chi$) was defined as:

$$\chi = \frac{\dot{n}_{in} - \dot{n}_{out}}{\dot{n}_{in}}$$

Where $\dot{n}_{in}$ and $\dot{n}_{out}$ are the inlet and outlet molar flow rates, respectively. The kinetic plots were obtained assuming a pseudo-first-order kinetics with respect to the analyte concentration, as expected for very low analyte concentrations.[97] Therefore, the mass-based reaction rates ($r$) were calculated according to ($Q_{tot}$ and $c_{in}$ are the total volumetric flow rate and inlet molar concentration, respectively):

$$r = \frac{Q_{tot} \cdot c_{in} \cdot \ln\left(\frac{1}{1-\chi}\right)}{m_{cat}}$$

### 4.6 Catalytic filter testing

The Pd/$SnO_2$ sensor was operated at 350 °C and connected in series with a packed bed of $CoCu_2O_3$ ($m_{cat}$ = 16 mg) nanoparticles heated to 170 °C. The as-realized detector (filter + sensor) was connected to the above mixing setup using the same gas standards. The total flow was kept constant at 150 mL $min^{-1}$ and the sensor was characterized as described above.

# Acknowledgements

This study was financially supported by the Swiss State Secretariat for Education, Research, and Innovation (SERI) under contract number MB22.00041 (ERC-STG-21 “HEALTHSENSE”). M.D. acknowledges Dr. Dmitry Chernyshov and staff of the Swiss-Norwegian Beamlines at ESRF for assistance with the high-temperature diffraction experiment. The authors acknowledge the Scientific Center for Optical and Electron Microscopy (ScopeM) of ETH Zurich for providing measuring time on their electron microscopes.

# Author contributions

M.D.: conceptualization, methodology, formal analysis, investigation, writing - original draft, writing - review & editing, visualization. T. E. A.-R. S.: methodology, formal analysis, investigation, writing - review & editing, visualization. E.C.: methodology, formal analysis, investigation. F.K.: methodology, formal analysis, investigation, writing - review & editing. A.T.G.: conceptualization, methodology, investigation, writing - review & editing, visualization, supervision, funding acquisition.

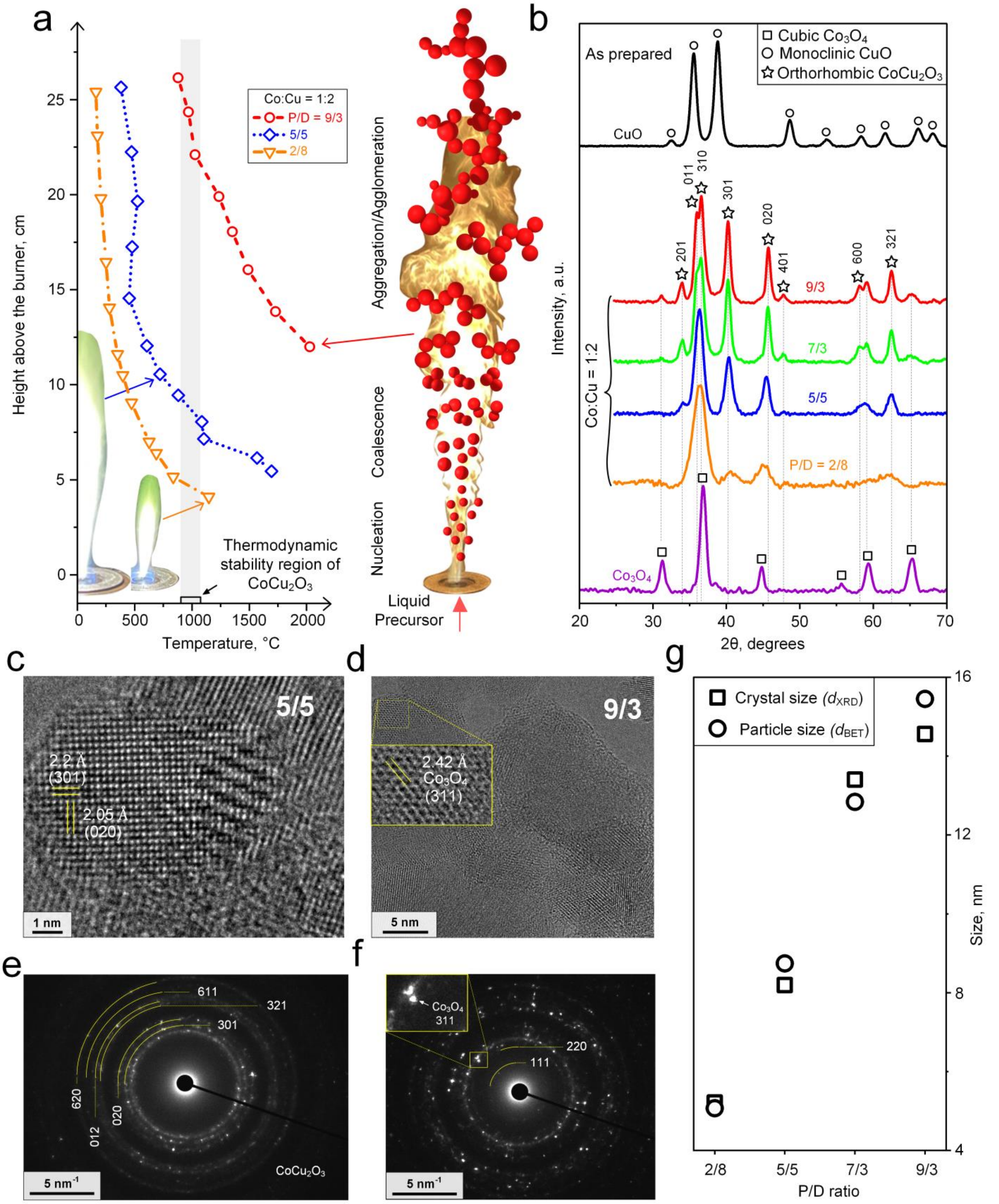

**Figure 1 | Combustion-aerosol synthesis of metastable $CoCu_2O_3$:** (a) Flame temperature profiles as a function of the height above the burner and different P/D ratios of 2/8 (triangles), 5/5 (diamonds) and 9/3 (circles), at a nominal Co:Cu ratio of 1:2. Corresponding pictures of 2/8, 5/5 and 9/3 flames, including the schematically illustrated particle formation pathway. (b) XRD patterns of as-prepared $Co_3O_4$, CuO and Co:Cu = 1:2 at P/D ratios of 2/8, 5/5, 7/3, 9/3. TEM images (c,d) and SAED patterns (e,f) of Co:Cu = 1:2 produced in 5/5 (c,e) and 9/3 (d,f) flames. (g) $CoCu_2O_3$ crystal ($d_{XRD}$) and BET-equivalent primary particle ($d_{BET}$) sizes at Co:Cu = 1:2 as a function of P/D.

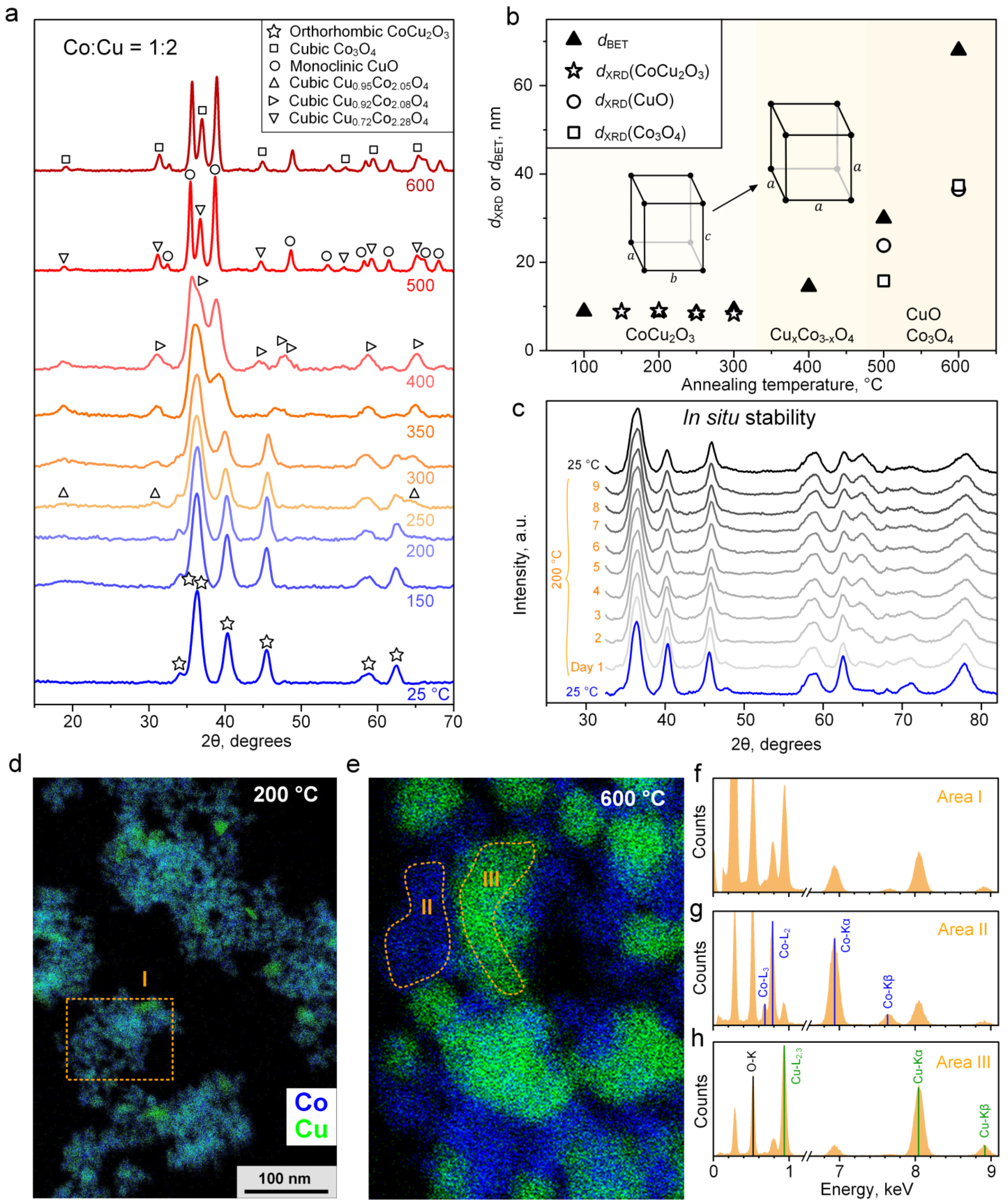

**Figure 2 | Thermal stability of $CoCu_2O_3$:** (a) XRD patterns of $CoCu_2O_3$ after a 5-hour treatment in air at increasing temperatures up to 600 °C. Indicated are the reference peaks of orthorhombic $CoCu_2O_3$ (stars), cubic $Co_3O_4$ (squares), monoclinic CuO (circles), cubic $Cu_{0.95}Co_{2.05}O_4$ (upward triangles), cubic $Cu_{0.92}Co_{2.08}O_4$ (right-pointing triangles) and cubic $Cu_{0.72}Co_{2.28}O_4$ (downward triangles). Note that diffractogram alignment is achieved by using an internal crystalline standard (see Methods). (b) $d_{XRD}$ (open symbols) and $d_{BET}$ (filled) after the treatment performed in (a). (c) XRD patterns of $CoCu_2O_3$ when kept over 9 days at 200 °C. Elemental maps (same scale bar) of powders annealed at (d) 200 and (e) 600 °C, showing the distribution of Cu (green) and Co (blue). (f) EDX spectrum of Area I in (d). Selected in (e) are Co-rich (Area II) and Cu-rich (III) zones, as confirmed by their individual EDX spectra in (g) and (h), respectively.

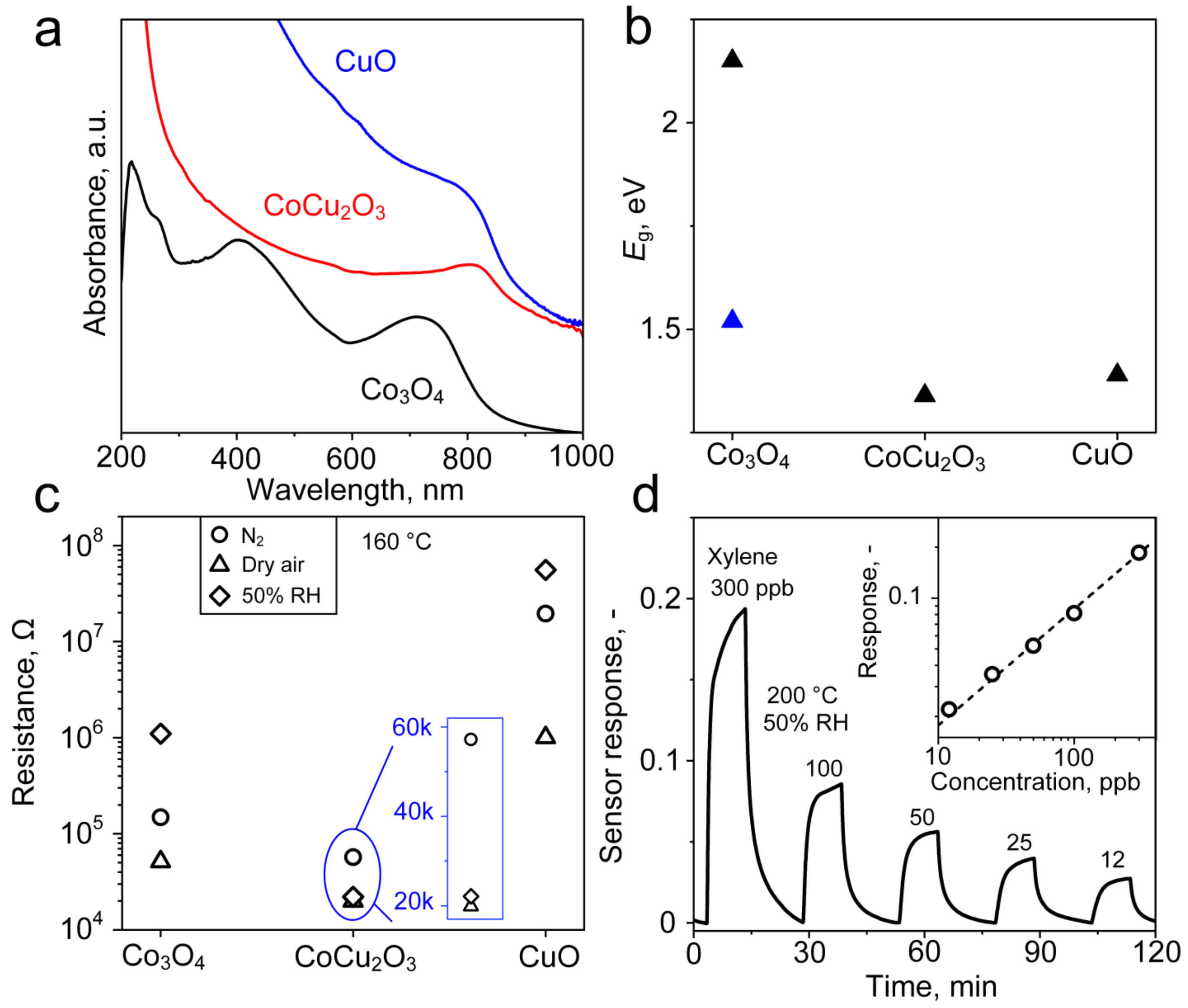

**Figure 3 | Optical, electric and chemoresistive properties:** (a) UV-Vis-NIR spectra of $Co_3O_4$, $CoCu_2O_3$ and CuO made in a flame with P/D = 5/5 after 5h treatment in air at 200 °C. (b) Corresponding estimated direct bandgap energies ($E_g$). (c) Electrical resistances of films of these nanoparticles when deposited directly from the flame-aerosol by thermophoresis onto substrates with interdigitated electrodes (see Methods). Electrical measurements are done in situ under different environments including inert $N_2$ (circles), dry air (triangles) and 50% humid air (diamonds). (d) Sensor response over time of $CoCu_2O_3$ upon exposure to 300 – 12 ppb xylene at 200 °C and under 50% RH with best fit (inset).

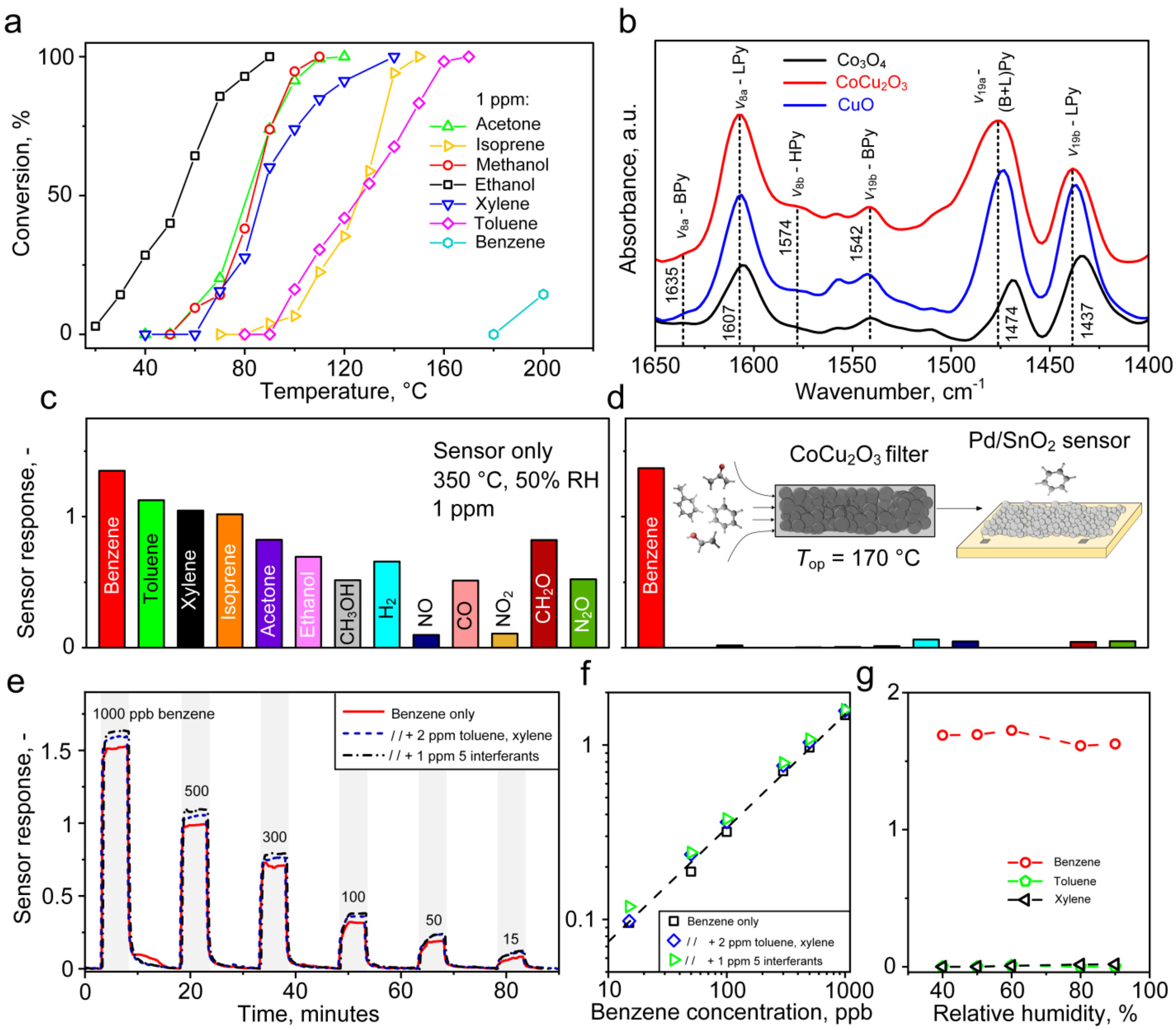

**Figure 4 | Catalytic activity and selective benzene detection:** (a) Conversion of acetone, isoprene, methanol, ethanol, xylene, toluene and benzene in air at 50% RH at 20 – 200 °C over $CoCu_2O_3$ nanoparticles. (b) Steady-state IR spectra of adsorbed pyridine at 200 °C for $Co_3O_4$, CuO and $CoCu_2O_3$. (c) Response to relevant VOCs for a Pd/$SnO_2$ sensor operated at 350 °C and 50% RH. (d) Responses adopting the detection scheme shown in the sketch inset and deploying the $CoCu_2O_3$ filter at operation temperature ($T_{op}$) of 170 °C. (e) Pd/$SnO_2$ – response transients upon exposure to decreasing concentrations of benzene (1000 – 15 ppb, red line) at 50% RH. The detector (filter + sensor) robustness to several-analyte interference is evaluated by co-feeding toluene and xylene concentrations of (each) 2 ppm (dashed line), and 1 ppm of up to 5 different interferants (dash-dotted line): toluene, xylene, methanol, ethanol and acetone. The corresponding responses are shown in (f). (g) Response of the filter-sensor system to benzene, toluene and xylene at different humidities in the range of 40 – 90% RH.

# Supplementary Information

# Metastable $CoCu_2O_3$

# for molecular sensing and catalysis

Matteo D'Andria,[1] Tiago Elias Abi-Ramia Silva,[1] Edoardo Consogno,[1]

Frank Krumeich[2] and Andreas T. Güntner[1,*]

[1] *Human-centered Sensing Laboratory, Department of Mechanical and Process Engineering, ETH Zurich, CH-8092 Zurich, Switzerland*
[2] *Laboratory of Inorganic Chemistry, Department of Chemistry and Applied Biosciences, ETH Zurich, CH-8093 Zurich, Switzerland*

* corresponding author: andregue@ethz.ch

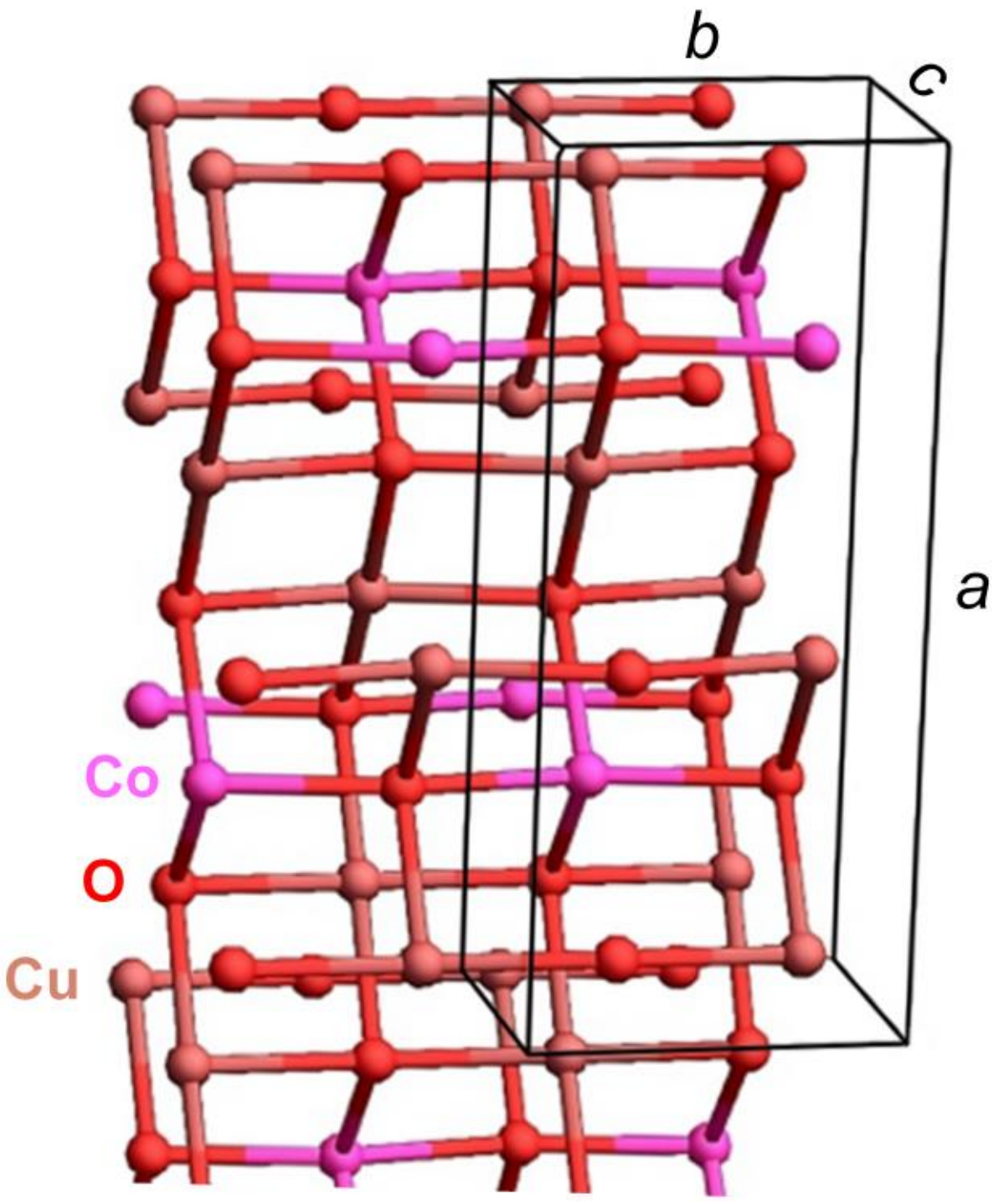

| | $a$ [Å] | $b$ [Å] | $c$ [Å] |
| --- | --- | --- | --- |
| Crystallographic reference | 9.409 | 3.980 | 3.198 |
| This work (5/5 flame) | 9.387 | 3.993 | 3.187 |

**Figure S1:** Orthorhombic $CoCu_2O_3$ lattice geometry. The table shows a comparison of the lattice parameters from this work, obtained by Rietveld refinement (P/D = 5/5), with the crystallographic reference [see main text].

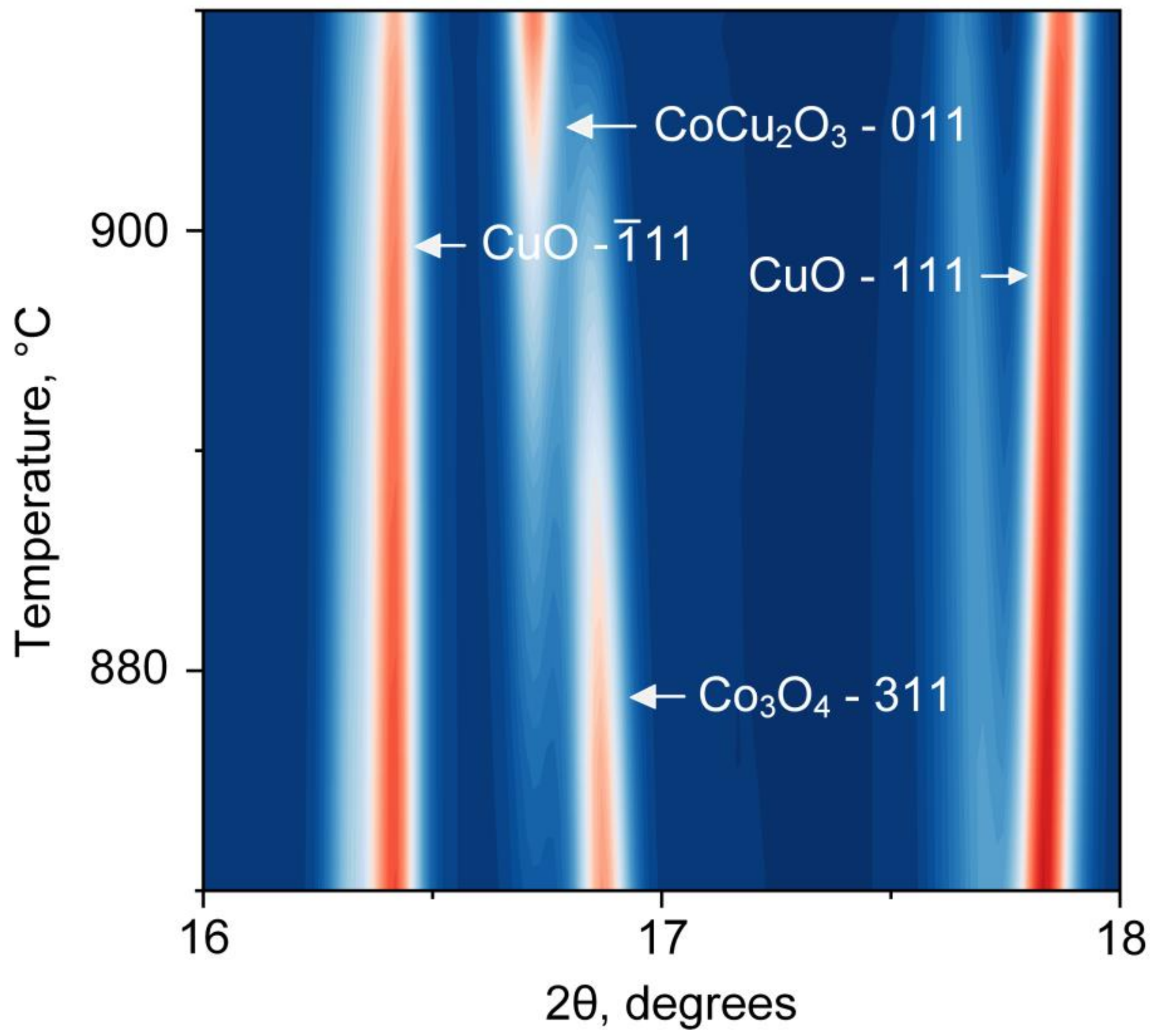

**Figure S2:** High-temperature XRD of Co:Cu = 2:1, showing the appearance of orthorhombic $CoCu_2O_3$ at high temperatures at $2\theta \approx 16.72°$. Please note that the $2\theta$ – range differs from Figure 1 and 2 due to the higher X-ray energy at the ESRF synchrotron facility ($\lambda$ = 0.72325 Å).

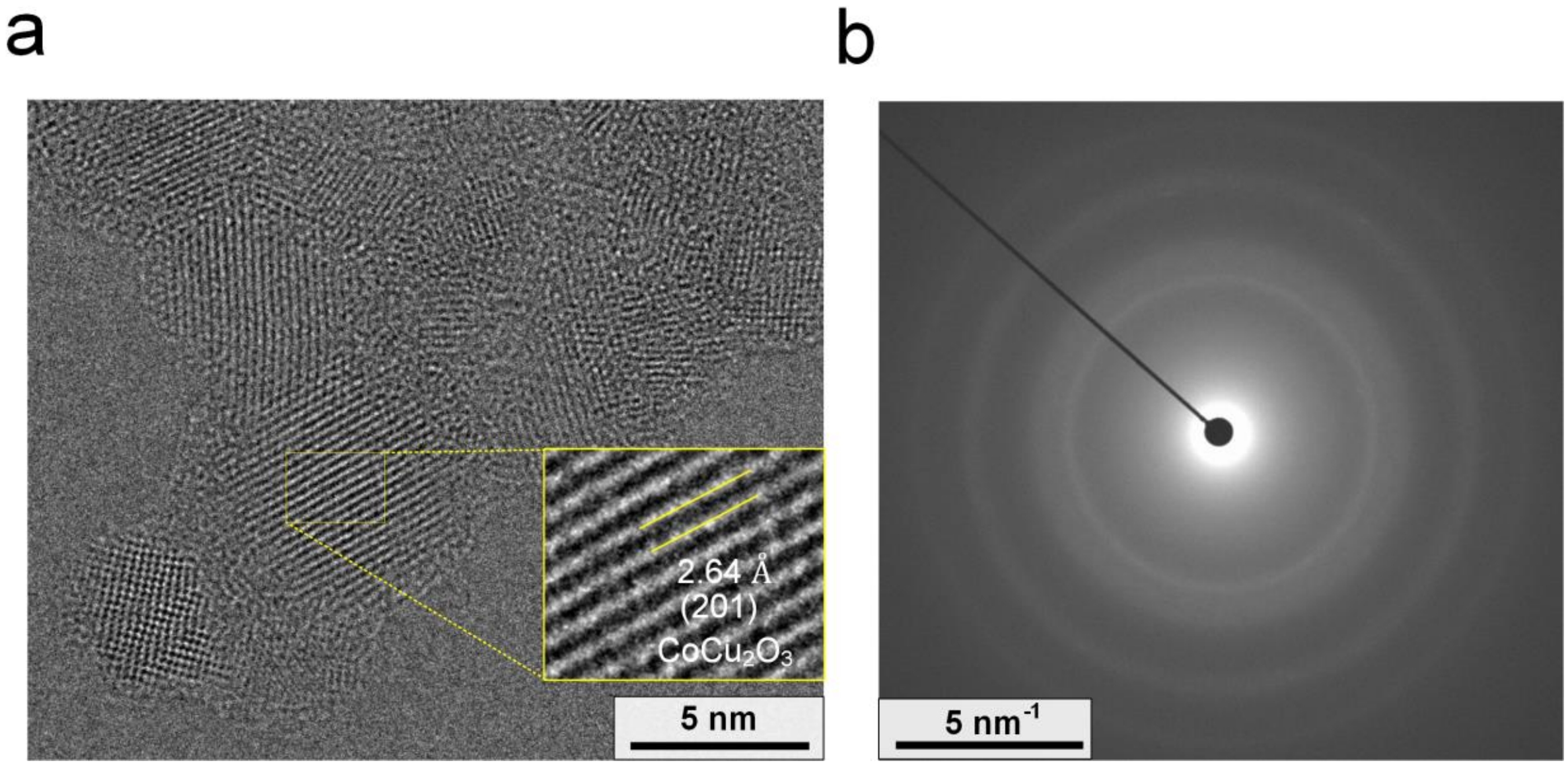

**Figure S3:** (a) TEM image of powders produced at Co:Cu = 1:2 at P/D = 2/8 with visible lattice fringes (see insets), as well as its (b) SAED pattern.

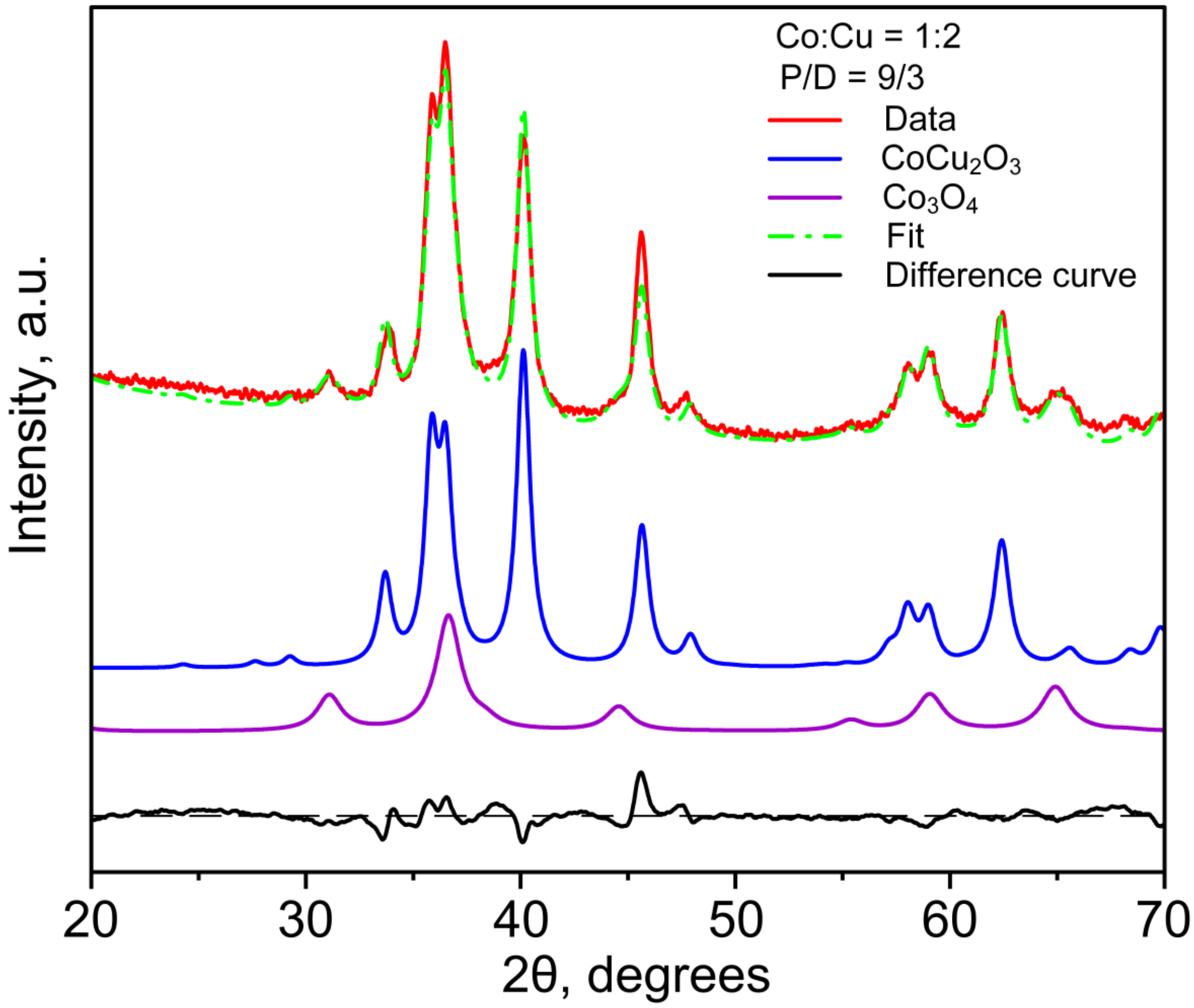

**Figure S4:** Rietveld refinement of XRD pattern of as-prepared 9/3 – powders.

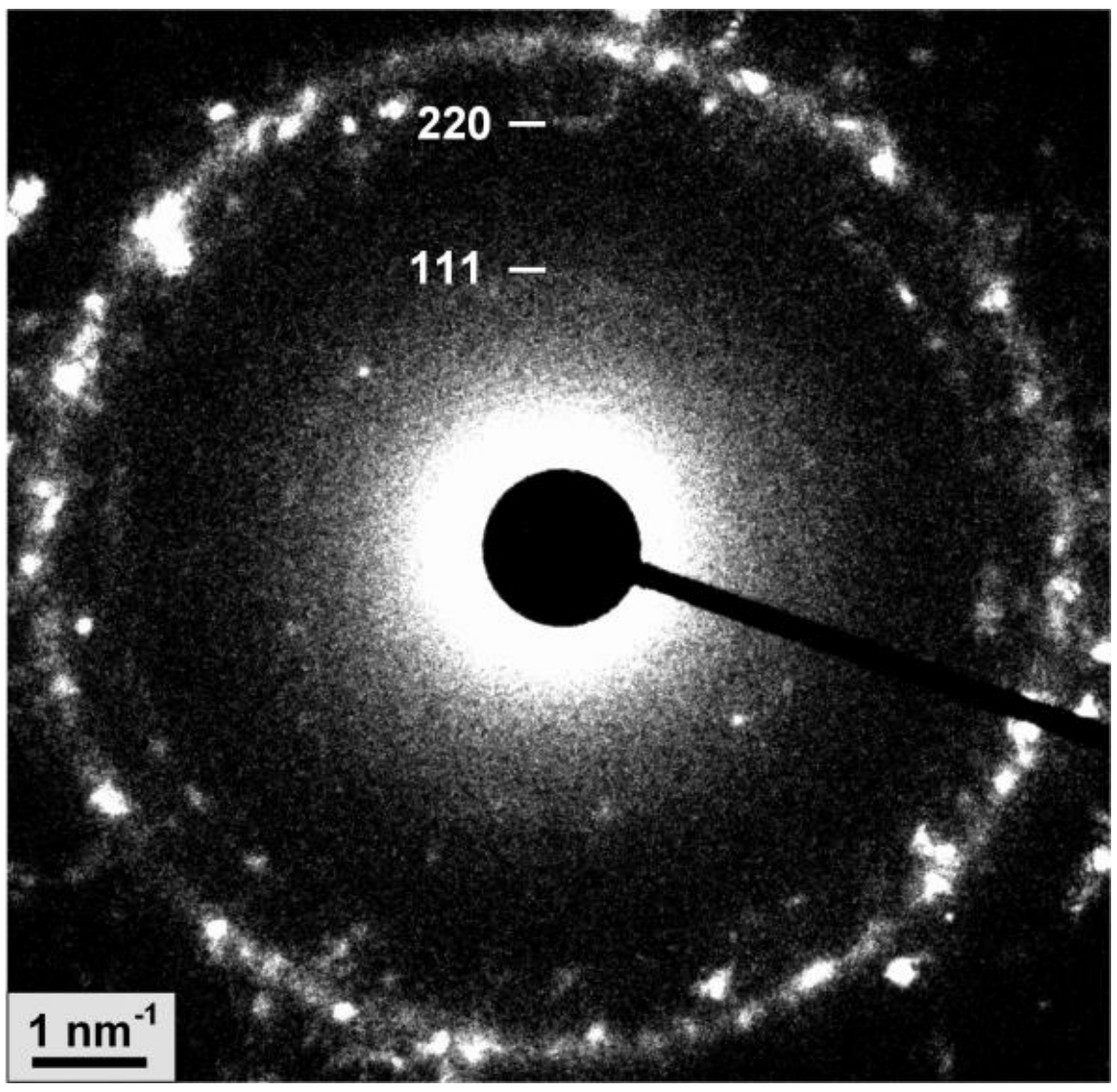

**Figure S5:** Electron diffraction pattern of Co:Cu = 1:2 at P/D = 9/3 showing the weak diffraction rings of $Co_3O_4$ indexed to 111 and 220.

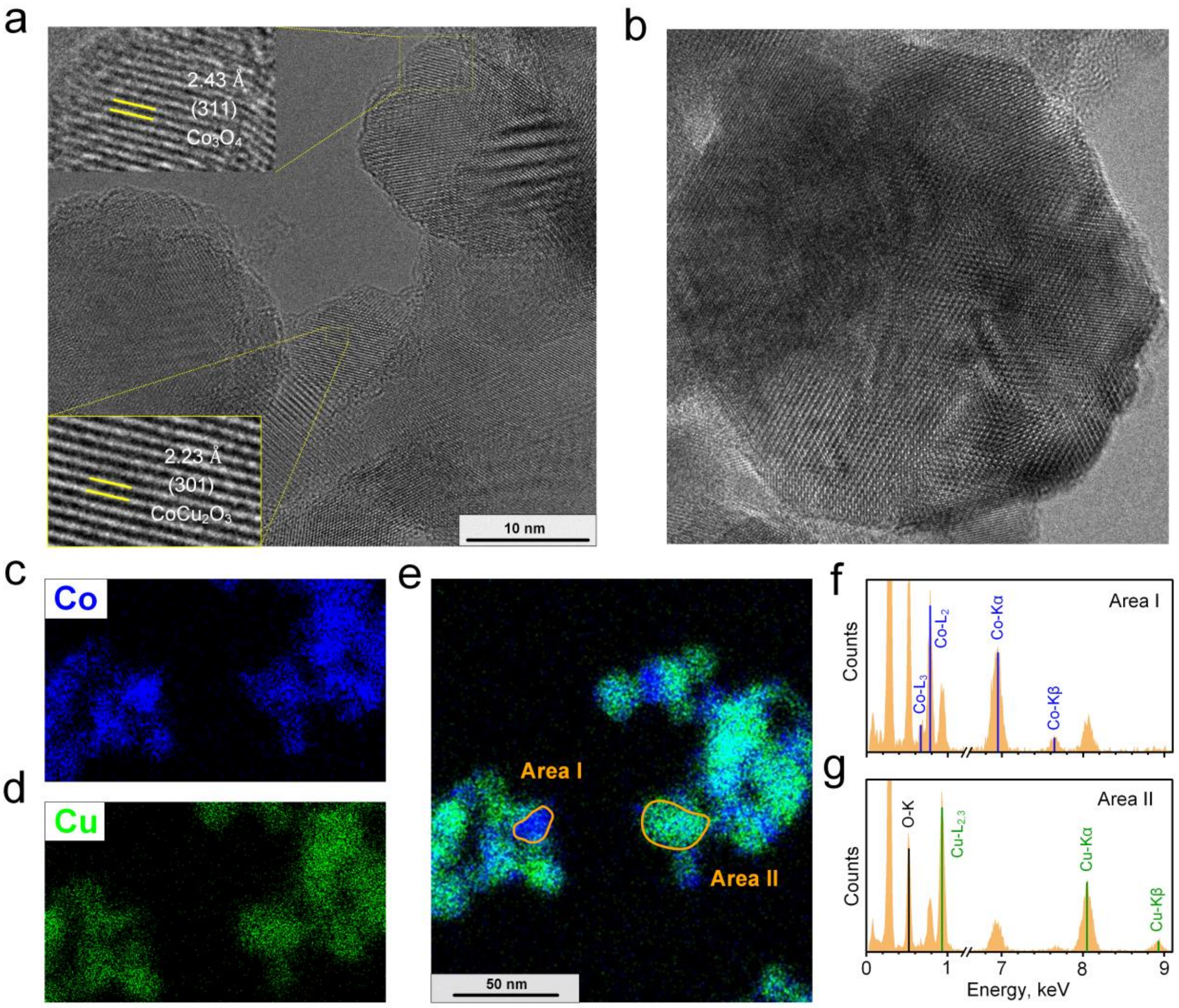

**Figure S6:** (a, b) TEM images of as-prepared $CuO$-$Co_3O_4$ powders (Co:Cu = 1:2) produced at P/D = 9/3. Some inhomogeneities can be observed, as confirmed by EDX elemental mapping of the same showing the distribution of (c) Co, (d) Cu and (e) the overlayed image. EDX spectra of selected areas in (e), showing the presence of (f) Co-rich (Area I) and (g) Cu-rich (II) zones.

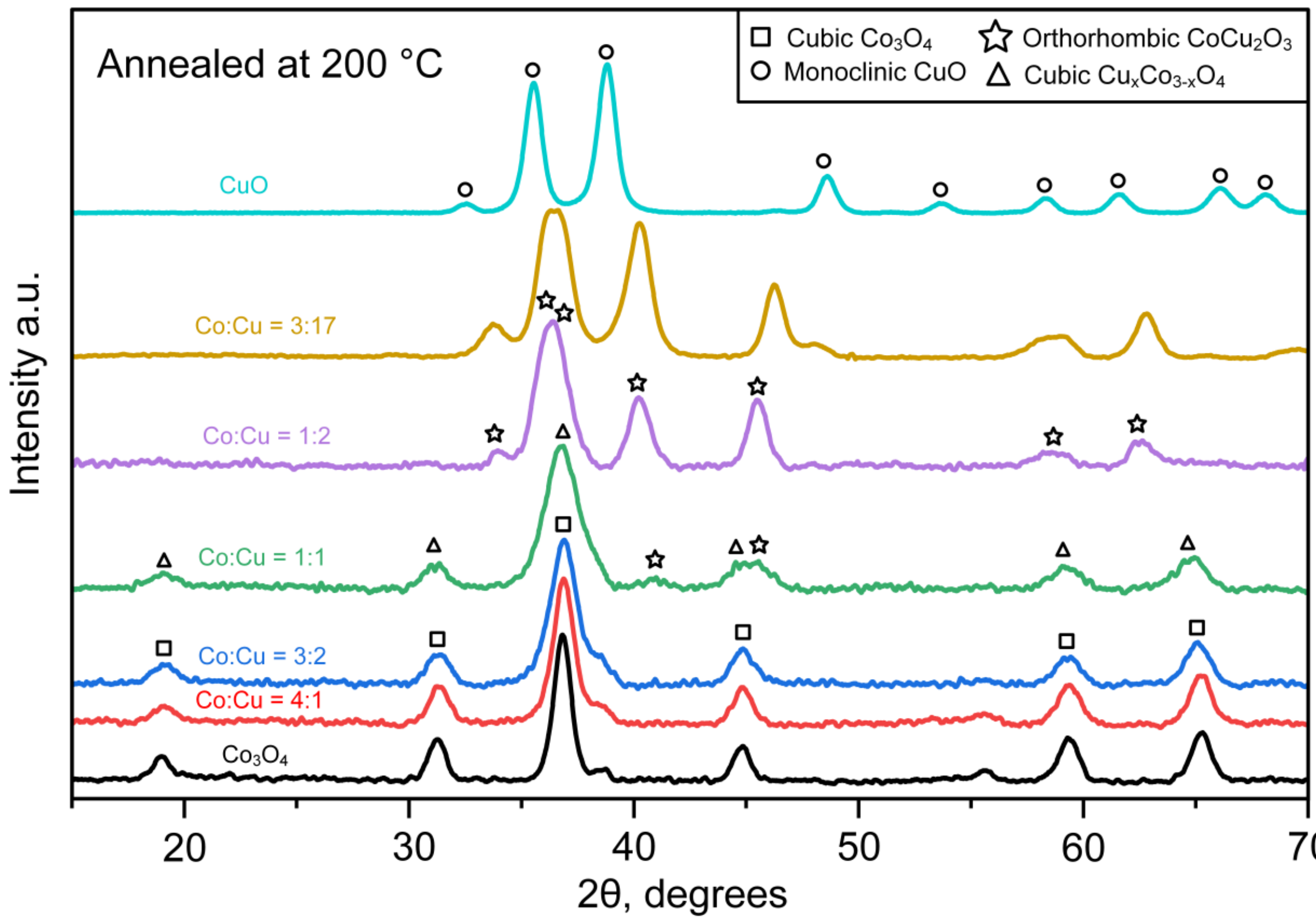

**Figure S7:** XRD patterns of annealed (200 °C) powders produced at different Co:Cu ratios. Indicated are the reference peaks of cubic $Co_3O_4$ (squares), monoclinic CuO (circles), cubic $Cu_xCo_{3-x}O_4$ (triangles) and orthorhombic $CoCu_2O_3$ (stars).

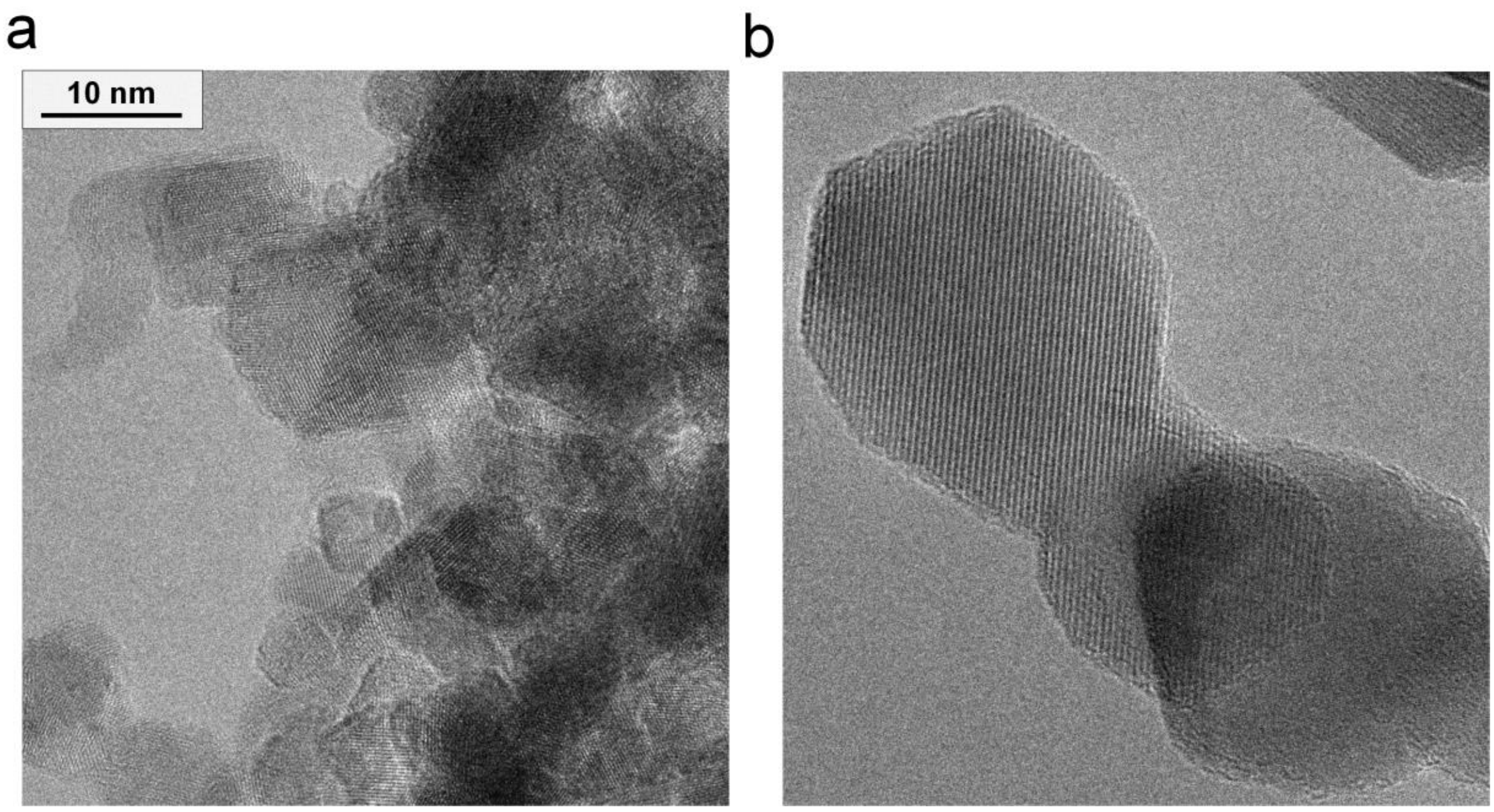

**Figure S8:** TEM images of Co:Cu = 1:2 annealed at (a) 200 °C and (b) 600 °C for 5 hours.

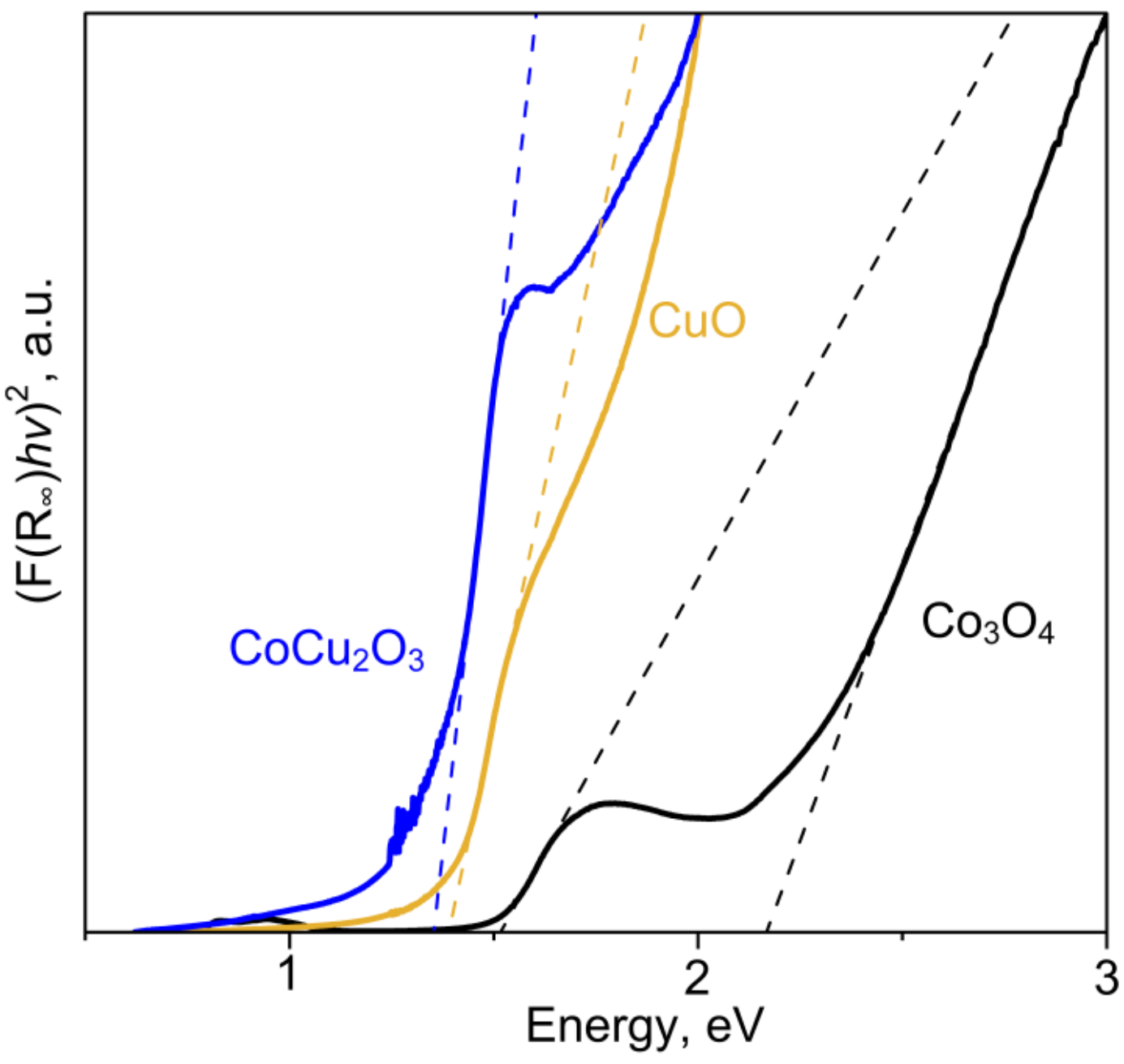

**Figure S9:** Tauc plots of $Co_3O_4$, $CoCu_2O_3$ and CuO.

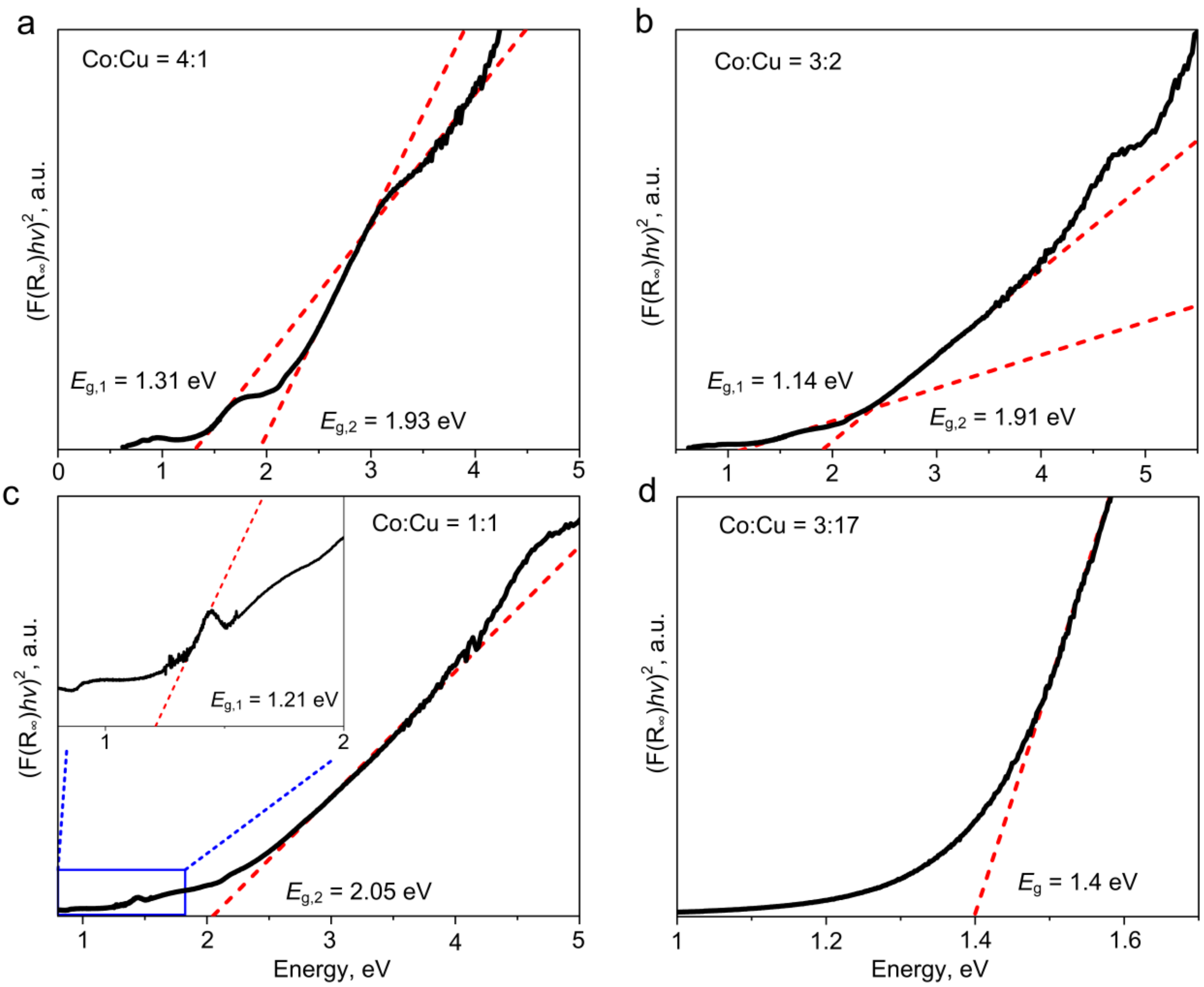

**Figure S10:** Tauc plots of CuO-$Co_3O_4$ powders prepared at Co:Cu ratios of (a) 4:1, (b) 3:2, (c) 1:1 and (d) 3:17. Indicated are the linear fits to extrapolate the direct electronic charge transfer energies.

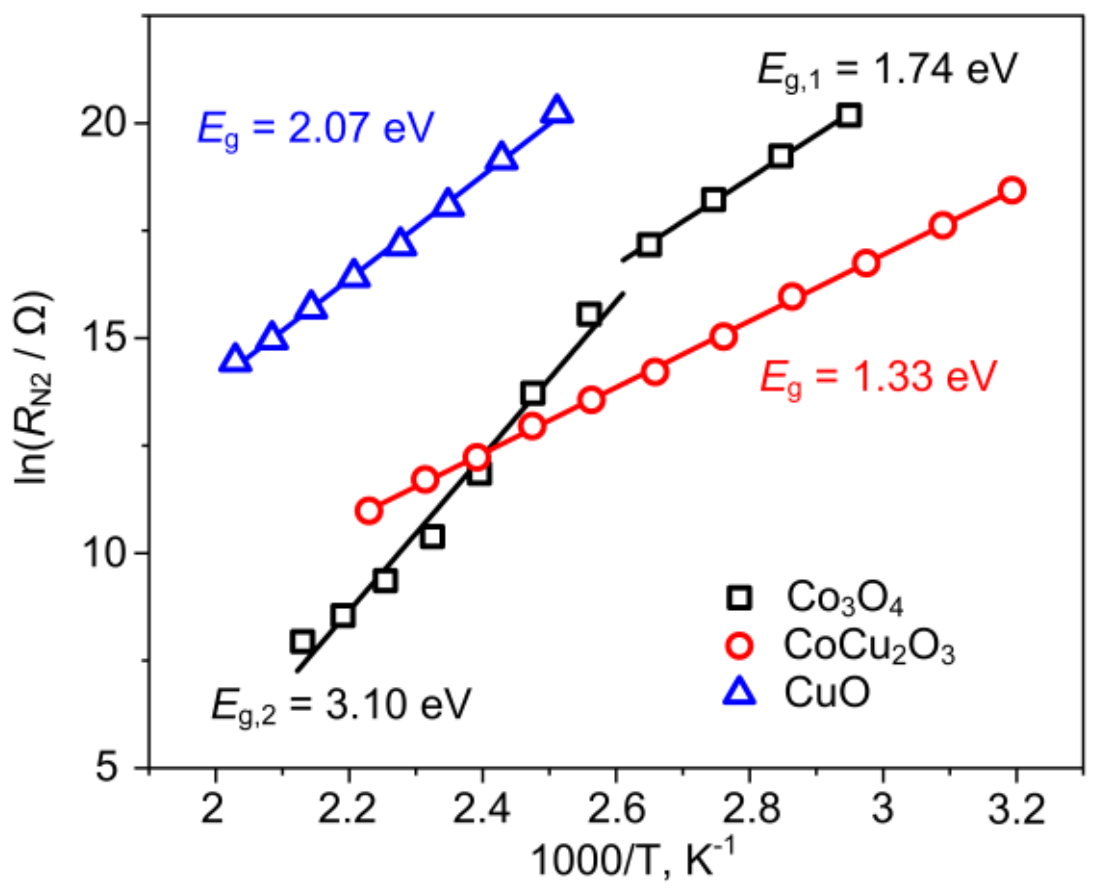

**Figure S11:** Natural logarithm of the ohmic resistance of $Co_3O_4$ (squares, black), $CoCu_2O_3$ (circles, red) and CuO (triangles, blue) vs. inverse temperature (with T ranging between 40 – 200 °C), measured under $N_2$. Indicated are the linear fits and corresponding values of thermal band-gap energies ($E_g$).

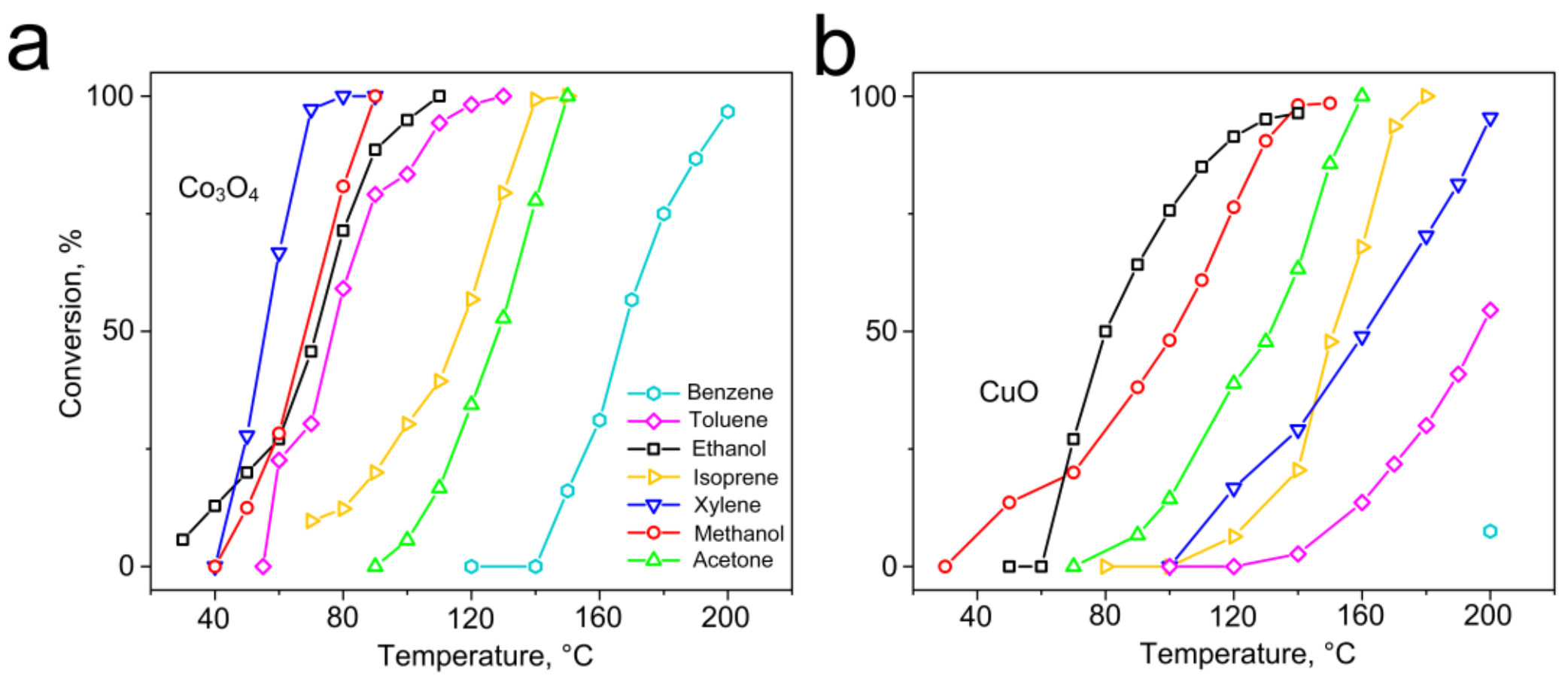

**Figure S12:** Catalytic conversion of 1 ppm acetone, isoprene, methanol, ethanol, xylene, toluene and benzene in air at 50% RH over (a) $Co_3O_4$ and (b) CuO nanoparticles annealed at 200 °C for 5 hours.

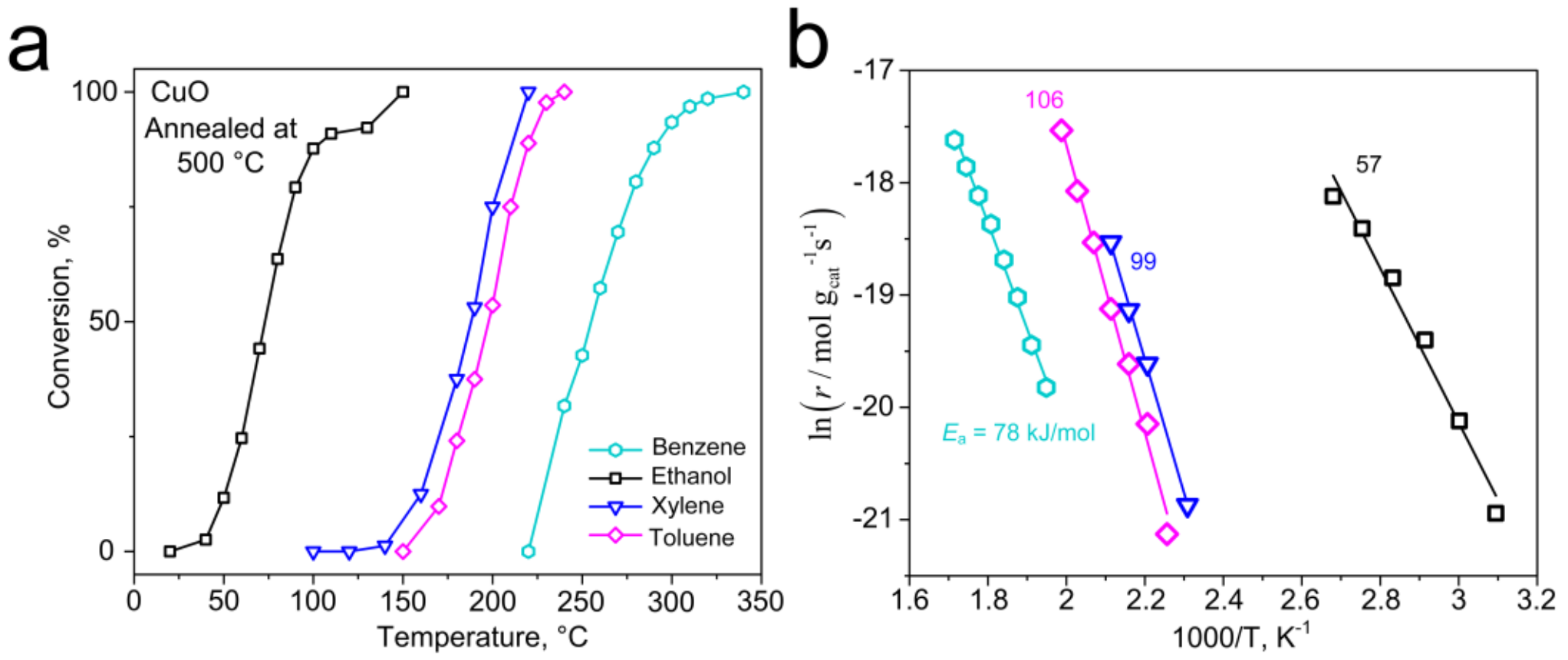

**Figure S13:** (a) Catalytic oxidation of 1 ppm benzene (hexagons), ethanol (squares), xylene (triangles) and toluene (diamonds) between 20 – 350 °C at 50% RH, over 16 mg of CuO nanoparticles, which were annealed at 500 °C for 5 hours. The kinetic plots, as well as the apparent activation energies, are shown in (b).

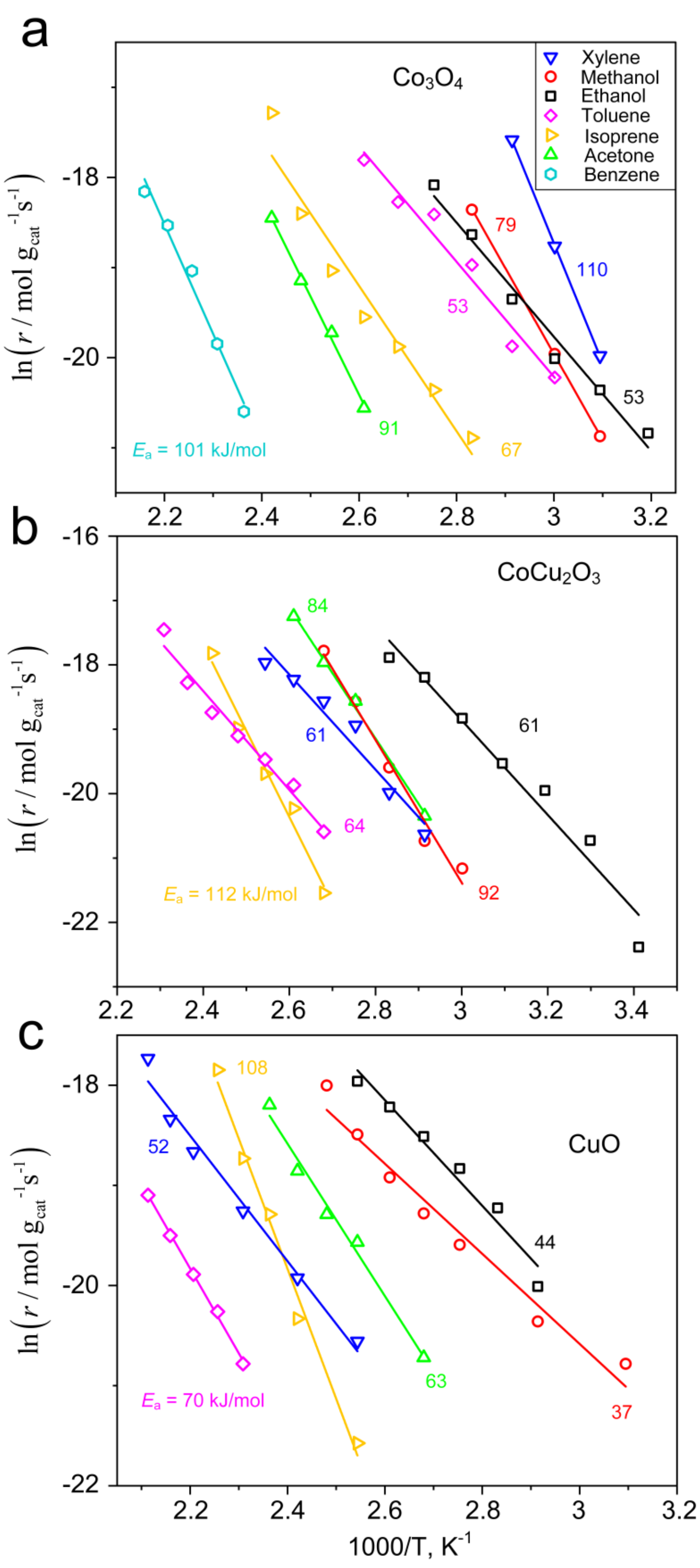

**Figure S14:** Arrhenius plots for catalytic oxidation over (a) $Co_3O_4$, (b) $CoCu_2O_3$ and (c) CuO.

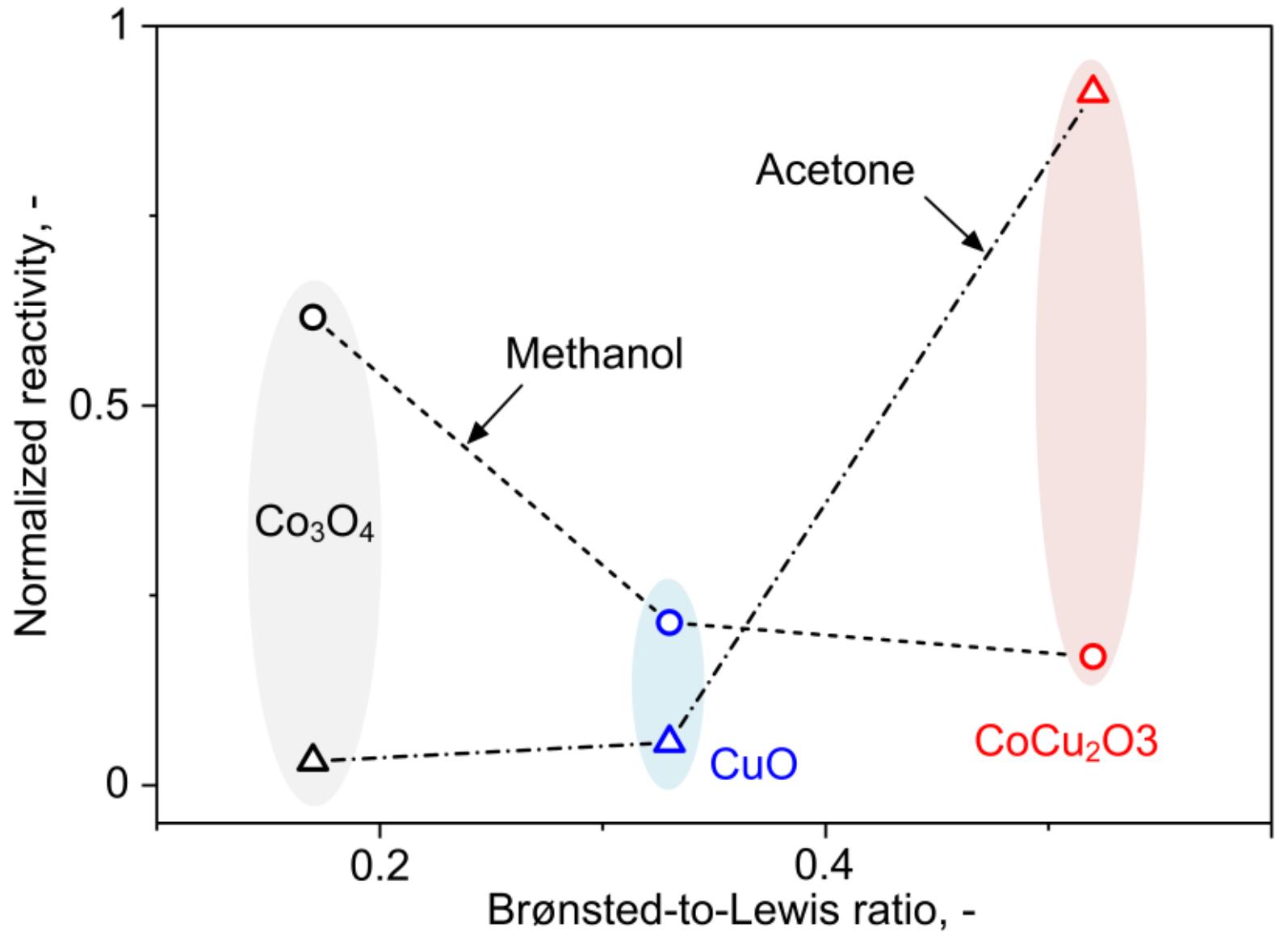

**Figure S15:** Normalized reaction rates of methanol and acetone as a function of the acid Brønsted-to-Lewis ratio, based on the $\nu_{19b}$ – intensities at 1541 cm$^{-1}$ (BPy) and 1440 cm$^{-1}$ (LPy). Reactivities of methanol and acetone were taken at 70 and 100 °C, respectively.

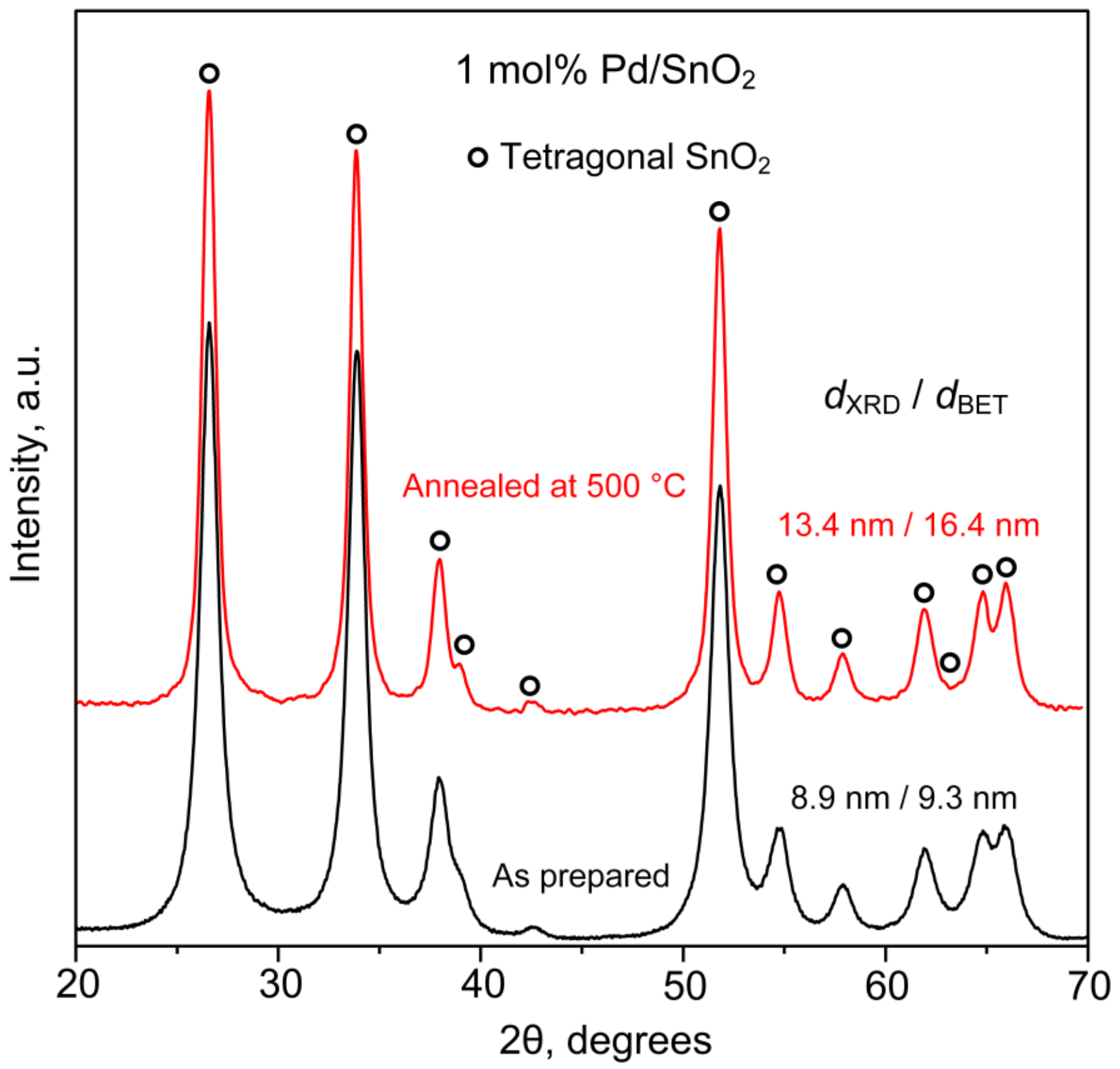

**Figure S16:** XRD patterns of as-prepared (black) and annealed (500 °C for 5 hours, red) 1 mol% $Pd/SnO_2$ nanoparticles. Indicated are the reference peaks of tetragonal $SnO_2$ (circles), as well as $d_{BET}$ and $d_{XRD}$.